\documentclass[12pt]{iopart}

\pdfoutput=1

\usepackage[utf8x]{inputenc}
\usepackage{subeqn}
\usepackage{wasysym}
\usepackage{graphicx}
\usepackage{tikz}
\usepackage{array}

\usepackage{mathrsfs}
\DeclareSymbolFontAlphabet{\mathrsfs}{rsfs}
\newcommand{\scri}{\mathrsfs{I}}

\newcommand{\CZ}{Z4c}
\newcommand{\K}{\Delta K}
\newcommand{\Te}{\bar \Theta}

\DeclareFontFamily{OT1}{rsfs}{}
\DeclareFontShape{OT1}{rsfs}{m}{n}{ <-7> rsfs5 <7-10> rsfs7 <10->
rsfs10}{} \DeclareMathAlphabet{\mycal}{OT1}{rsfs}{m}{n}

\begin{document}

\title[Spherically symmetric unconstrained hyperboloidal evolution]{Spherical symmetry as a test case for unconstrained hyperboloidal evolution}

\author{Alex Vañó-Viñuales$^1$, Sascha Husa$^1$ and David Hilditch$^2$}

\address{$^1$ Universitat de les Illes Balears and Institut d'Estudis Espacials de Catalunya, Cra. de Valldemossa km.7.5, 07122 Palma de Mallorca, Spain}
\address{$^2$ Theoretisch-Physikalisches Institut, Friedrich-Schiller-Universität Jena, Max-Wien-Platz 1, 07743 Jena, Germany}
\ead{alex.vano@uib.es}
\begin{abstract}

We consider the hyperboloidal initial value problem for the Einstein equations in numerical relativity, motivated by the goal to evolve radiating compact objects such as black hole binaries with a numerical grid that includes null infinity. Unconstrained evolution schemes promise optimal efficiency, but are difficult to regularize at null infinity, where the compactified Einstein equations are formally singular. In this work we treat the spherically symmetric case, which already poses nontrivial problems and constitutes an important first step. We have carried out stable numerical evolutions with the generalized BSSN and Z4 equations coupled to a scalar field. The crucial ingredients have been to find an appropriate evolution equation for the lapse function and to adapt constraint damping terms to handle null infinity. 

\end{abstract}

\maketitle

\section{Introduction}

It is common in general relativity to model astrophysical processes with 
spacetimes that are asymptotically flat along the null geodesics that ``escape" from the source. In particular, it is known that if the duration of a radiation signal from an astrophysical event is much shorter than the travel time from the source to the observer, then the observer can be appropriately idealized as being located at future null infinity \cite{Barack:1998bv,Leaver1986,PhysRevD.34.384}.  Computing observable quantities at null infinity allows to compute the energy loss and gravitational wave signal without ambiguities - which seems not possible at finite distance to the source, essentially due to the equivalence principle. 

In numerical simulations of astrophysically interesting situations, such as compact binary coalescence, it is however still customary to extract the gravitational wave signal from the source at a series of finite radii, and extrapolate to the result at null infinity, see \cite{Hinder:2013oqa} for a recent comparison of results from different codes.
Solving radiation problems employing a foliation of spacetime that reaches null infinity not only removes ambiguities, it can also be more efficient: on a spacelike slice that reaches spacelike infinity an infinite number of wave cycles would need to be resolved. Looking inward  along a null slice toward a radiation source from null infinity, the signal is ``in phase'' with the source, thus only a single cycle needs to be resolved. 
Extrapolation from a series of finite extraction radii and the use of spacelike hyperboloidal slices blur this distinction, but it still appears clear that for efficiently computing radiation signals from astrophysical events, it is desirable to numerically handle domains of infinite physical size and compute radiation signals directly at null infinity. 

Based on the work of Penrose \cite{PhysRevLett.10.66,Penrose:1965am}, singular conformal rescalings 
have become a standard tool to describe asymptotics in general relativity in terms of local differential geometry on a compactified unphysical spacetime with metric $\bar g_{ab}$.  Under these Penrose-like rescalings the physical metric $\tilde g_{ab}$ is related to the unphysical one as
\begin{equation}\label{conftrafog} \bar g_{ab} \equiv\Omega^2\tilde g_{ab} ,  \end{equation} 
and the boundary of the physical spacetime at infinite distance is then characterized by the vanishing of the conformal factor, $\Omega=0$. However, this comes at the price of singularities in the Einstein equations rewritten in terms of $\bar g_{ab}$ and $\Omega$. For numerical evolutions, these singularities need to be regularized analytically or treated with extreme care. Due to the complexity of the Einstein equations this has proven difficult for strong field astrophysical systems.

The problem is significantly simplified in the characteristic approach, where one evolves along null slices \cite{PhysRevLett.80.3915,Winicour2009LRR}. In spherical symmetry this approach has become very popular, see e.g. \cite{1983RSPSA.386..373C,1999.Christodoulou,PhysRevD.36.3575,Gomez199211,PhysRevD.51.5558, Husa:2000kr}, and conformal compactification is essentially trivial. In general situations such slices are however prone to produce caustics. A very successful pragmatic approach has been characteristic extraction, where data for the characteristic evolution are specified on a null cone and a timelike tube at sufficient separation from the source. These data are determined by a standard numerical evolution of a finite but sufficiently large region of spacetime, such that its outer boundaries are causally disconnected from the timelike tube where characteristic data are extracted \cite{Reisswig:2009us}. This approach has in particular been used to study the accuracy of extrapolation from finite radii \cite{Reisswig:2009rx,Taylor:2013zia}.

A more flexible alternative is the hyperboloidal initial value problem, pioneered by Friedrich \cite{friedrich1983,lrr-2004-1,Friedrich86} where one evolves spacelike slices that reach null infinity: being smooth and the absence of matching interfaces or abrupt changes of the causality structure are their main advantages. 
In this case, it has been possible to completely regularize the conformally rescaled Einstein equations without assuming a particular choice of coordinates, however at the price of a significant increase in evolution variables and complexity, which makes the evolution equations prone to continuum instabilities \cite{Husa:2005ns,Husa:2002zc}, instabilities excited by numerical perturbations that satisfy the continuum equations. 
In the context of numerical simulations of astrophysical systems however, there is a preferred class of coordinate systems at null infinity, Bondi coordinates (see e.g. \cite{Wald} for a textbook treatment), which recover the simplicity of Minkowski space. One can hope to construct a simple evolution system which is beneficial for numerical evolutions by only regularizing the conformally rescaled Einstein equations after adopting the Bondi gauge at null infinity, and exploiting all its simplifications. A way to proceed is to adopt an elliptic-hyperbolic evolution system, where boundary conditions for the coordinate gauge (e.g. for lapse and shift vector) can be specified for the elliptic part \cite{Andersson:springer,Rinne:2009qx,Rinne:2013qc}. 
This has been a very successful approach in numerical simulations, however the numerical solution of elliptic equations is computationally expensive and the specification of inner boundaries subtle in the presence of horizons. 
A purely hyperbolic evolution system (a special case of unconstained or free evolution, where the solution is constructed by integrating the hyperbolic differential equations in time and the constraint equations are only used to monitor the behaviour of the solution) would be desirable, and this is the approach we want to take in the present work. We restrict ourselves to the already nontrivial case of spherical symmetry as a test problem, to see whether our goal of robust numerical evolutions can be achieved.

Our approach can be summarized as follows:
In section \ref{eqshyp} we start from the generalized Baumgarte-Shapiro-Shibata-Nakamura (GBSSN) \cite{PhysRevD.52.5428,Baumgarte:1998te,Brown:2007nt} and conformal Z4 \cite{bona-2003-67,Alic:2011gg,Bernuzzi:2009ex,Sanchis-Gual:2014nha} equations, which are customary in numerical relativity, and rewrite them in terms of finite conformally rescaled variables, following Zengino\u{g}lu's approach of working with a time independent conformal factor \cite{Zenginoglu:2007jw,Zenginoglu:2008wc,Zenginoglu:2008pw,Zenginoglu:2008uc}. We describe the derivation of our initial data for Minkowski spacetime on the hyperboloidal foliation in section \ref{inihyp}. 
In section \ref{gaugecond} we fix null infinity to a fixed value of the radial coordinate (scri-fixing \cite{PhysRevD.58.064003}) by specifying appropriate gauge conditions. In contrast to  Zengino\u{g}lu's work \cite{Zenginoglu:2007jw,Zenginoglu:2008wc,Zenginoglu:2008pw,Zenginoglu:2008uc}
we will however not make use of the preferred conformal gauge, which would correspond to using a Bondi time parametrization at null infinity.
After analyzing hyperbolicity of our systems of evolution equations in section \ref{hypanal}, we describe our numerical finite difference methods in section \ref{nummeth}.  Continuum instabilities in the evolution equations are studied with analytical and numerical methods employing different subsets of the full equation system, and cured by modifying our choice of evolution variables and gauge conditions to avoid such instabilities, as explained in detail in section \ref{regularizingsec}.
We demonstrate the robust numerical evolution and convergence of strong ``gauge wave'' signals  and  self-gravitating scalar fields in section \ref{resu}.

We employ the following conventions and notation for the metrics: the 4-dimensional physical metric is denoted as $\tilde g$, the 4d-conformal metric as $\bar g$, the 3d conformal metric (induced by $\bar g$) as $\bar \gamma$, the 3d twice conformal metric $\gamma$ and the 3d twice conformal background metric $\hat \gamma$.

\section{Conformally compactified equations on hyperboloidal foliations} \label{eqshyp}

\subsection{Conformally compactified equations}

The Einstein equations for a metric $\tilde g_{ab}$ including the constraint propagation terms of the Z4 formalism \cite{bona-2003-67,Bona:2003qn} and their damping terms \cite{Gundlach:2005eh} proportional to the timelike normal vector $\tilde n^a$ (with the parameters $\kappa_1$ and $\kappa_2$ chosen empirically) are given by 
\begin{equation}\label{einsteinphyseq}
G[\tilde g]_{ab}+2\tilde\nabla_{(a} Z_{b)} -\tilde g_{ab}\tilde\nabla^cZ_c - \kappa_1\left(2\,\tilde n_{(a}Z_{b)}+\kappa_2\,\tilde g_{ab}\,\tilde n^cZ_c\right) =8\pi T_{ab}. 
\end{equation}
The Einstein equations are satisfied when the field $Z_a$ vanishes.
Written in terms of the conformally rescaled metric $\bar g_{ab}$ as defined in \eref{conftrafog} and the rescaled normal vector $\bar n^a\equiv\tilde n^a/\Omega$ the Einstein equations become 
\begin{equation}\label{einsteineq}
\fl G[\tilde g]_{ab}+2\bar\nabla_{(a} Z_{b)} -\bar g_{ab}\bar\nabla^cZ_c + \frac{4}{\Omega} Z_{(a}\bar\nabla_{b)}\Omega - \frac{\kappa_1}{\Omega}\left(2\,\bar n_{(a}Z_{b)}+\kappa_2\,\bar g_{ab}\,\bar n^cZ_c\right)=8\pi T_{ab}  . 
\end{equation}
The Einstein tensor of the physical metric, $G[\tilde g]$, is related to that of the conformal metric, $G[\bar g]$, as
\begin{equation}\label{einsteinten}
G[\tilde g]_{ab} = G[\bar g]_{ab} + {2\over\Omega}(\bar\nabla_a\bar\nabla_b\Omega-\bar g_{ab}\bar\Box \Omega) + {3\over\Omega^2}\bar g_{ab}(\bar\nabla_c\Omega)(\bar\nabla^c\Omega) .
\end{equation}
The conformal factor terms (the two last terms in \eref{einsteinten}) are divergence-free without requiring any additional conditions on $\Omega$ and also attain a regular limit at $\scri^+$ for the preferred conformal gauge choice, as shown in \cite{Zenginoglu:2007it,Zenginoglu:2008pw}, see Sec.~\ref{gaugecond} for more details.
Indices are raised and lowered with the conformal metric $\bar g_{ab}$, its covariant derivative is denoted by $\bar\nabla$ and $\bar\Box\equiv\bar g^{ab}\bar\nabla_a\bar\nabla_b$.

Here we have introduced the Z4 quantities in the physical Einstein equations \eref{einsteinphyseq}, but in principle they could also be added at the level of the conformal metric equations. In this case, the fourth term in \eref{einsteineq} would not appear and the Z4 damping term would not be divided by the conformal factor $\Omega$. The reason why we chose to use \eref{einsteineq} as it is presented above is that it automatically gives us the damping terms that make the equations well-behaved, see subsection \ref{constraintdamping}.

In our work we will choose the energy-momentum tensor $T_{ab}$ to describe a massless scalar field $\Phi$:
\begin{equation}
T_{ab} = \tilde \nabla_a\Phi\tilde\nabla_b\Phi-\case{1}{2}\tilde g_{ab}\,(\tilde\nabla_c\Phi)(\tilde\nabla^c\Phi) \ = \bar \nabla_a\Phi\bar\nabla_b\Phi-\case{1}{2}\bar g_{ab}\,(\bar\nabla_c\Phi)(\bar\nabla^c\Phi)   , 
\end{equation}
with field equation
\begin{equation}\label{4dkgfphys}
\tilde\Box\Phi=0  ,
\end{equation} 
which in the conformally rescaled geometry becomes 
\begin{equation}\label{4dkgf}
\bar\Box\Phi-2\,\Omega^{-1}(\bar\nabla_a\Phi)(\bar\nabla^a\Omega)=0\,. 
\end{equation}

\subsubsection{3+1 decomposition of the Einstein equations}\label{sub3+1}

We perform a standard 3+1 decomposition of the Einstein equations \eref{einsteineq} in terms of the timelike unit normal vector to the spacelike supersurfaces, $\bar n^a$, and the induced spatial metric $\bar\gamma_{ab} = \bar g_{ab} +\bar n_a \bar n_b$. We also introduce the conformal extrinsic curvature $\bar K_{ab}$ as 
\begin{equation}
\bar K_{ab} = -\case{1}{2}\mathcal{L}_{\bar n}\bar\gamma_{ab} =-\case{1}{2\alpha}\partial_\perp\bar\gamma_{ab},
\end{equation}
which is related to the physical extrinsic curvature, $\tilde K_{ab} = -\case{1}{2}\mathcal{L}_{\tilde n}\tilde\gamma_{ab} =-\case{\Omega}{2\alpha}\partial_\perp\tilde\gamma_{ab}$, as
\begin{equation}
\tilde K_{ab} = \frac{1}{\Omega}\bar K_{ab} +\frac{\bar \gamma_{ab}}{\alpha}\frac{\partial_\perp\Omega}{\Omega^2}  . 
\end{equation}

The energy-momentum terms are decomposed using $\rho=\bar n^a\bar n^bT_{ab}$, $J^a=-\bar\gamma^{ab}\bar n^cT_{bc}$ and $S_{ab}=\bar\gamma_a^c\,\bar\gamma_b^d\,T_{cd}$, with $S=S_a^a$. The variable $\Theta$ is introduced as the perpendicular projection of $Z_a$, that is $\Theta=-\bar n^a Z_a$, and from now on we will refer to $Z_a$ as the 3-dimensional spatial projection of the original Z4 quantity. 
The full 3+1 decomposed equations are presented in \ref{app3+1}, where we also introduce a parameter $C_{Z4c}$ to handle alternative treatments in the literature of certain non-principal part Z4 terms (which vanish when the constraints are satisfied): For $C_{Z4c} = 1$ all Z4 terms are kept as in the CCZ4 formulation \cite{Alic:2011gg}, while for  $C_{Z4c} = 0$ some non-principal part terms are dropped as in the  Z4c formulation \cite{Bernuzzi:2009ex,Weyhausen:2011cg}.

\subsubsection{GBSSN and \CZ{} systems}\label{subconf}

For the GBSSN formulation we will mainly follow \cite{Brown:2009dd}. The Z4 variables $\Theta$ and $Z_a$ will be treated as in \cite{Bernuzzi:2009ex,Weyhausen:2011cg}.  For the numerical treatment of roughly spherical compact objects such as black holes or neutron stars it has been found beneficial to factor out a conformal factor from the 3-dimensional geometry. Following standard procedure
we define the 3-dimensional (twice-)conformal metric $\gamma_{ab}$ and the conformal factor $\chi$ that relates it to $\bar \gamma_{ab}$, as well as the conformal trace-free part of the extrinsic curvature $A_{ab}$ and $K$, its trace $\bar K=\bar\gamma^{ab}\bar K_{ab}$ mixed with $\Theta$: 
\begin{equation}
\gamma_{ab}=\chi\bar \gamma_{ab}, \quad A_{ab}=\chi\left(\bar K_{ab}-\frac{1}{3}\bar\gamma_{ab}\bar K\right), \quad K=\bar K -2\Theta.
\end{equation}
To achieve hyperbolicity we add an evolution equation for a connection-type quantity. We first define the quantity $\Delta\Gamma^a$, the difference between two contracted Christoffel symbols, one related to $\gamma_{ab}$ and the other one to a time-independent background metric $\hat\gamma_{ab}$. The actual evolution variable we use is then $\Lambda^a$, which is defined from $\Delta\Gamma^a$ and the Z4-quantity $Z_a$:
\begin{equation}\label{dglz}
\Delta\Gamma^a=\gamma^{bc}\Delta\Gamma^a_{bc}=
\gamma^{bc}\left(\Gamma^a_{bc}-\hat\Gamma^a_{bc}\right), \quad \Lambda^a=\Delta\Gamma^a+2\gamma^{ab}Z_b.
\end{equation}
The complete system of evolution equations and constraints is listed in \ref{apptens}. 
We can easily choose the \CZ{} system or the GBSSN one simply by evolving the variable $\Theta$ and substituting $Z_{a}$ by $\Lambda^a$ as in \eref{dglz} or by setting the two Z4 quantities to zero, respectively.

\subsubsection{GBSSN and \CZ{} in spherical symmetry}\label{subspher}

We will now restrict to the spherically symmetric case. All evolution variables will depend only on the radial coordinate $r$ and on the time coordinate $t$, and we will suppress this dependence in our notation for simplicity.
%
We write the metric in terms of two independent quantities $\gamma_{rr}$ and $\gamma_{\theta\theta}$. Having previously introduced the spatial conformal factor $\chi$, we could now eliminate either $\gamma_{rr}$ or $\gamma_{\theta\theta}$ by fixing the determinant of $\gamma_{ab}$ as convenient. For $A_{ab}$ we explicitly impose its trace-freeness \cite{Brown:2007nt}: 
\begin{equation}
\fl  { \gamma}_{ij} = \left( \begin{array}{ccc} { \gamma}_{rr} & 0 & 0 \\
                                         0 & r^2{ \gamma}_{\theta\theta} & 0 \\
                                         0 & 0 & r^2{ \gamma}_{\theta\theta}\sin^2{\theta} 
                            \end{array}  \right) , \ \qquad
  { A}_{ij} = { A}_{rr} \left( \begin{array}{ccc} 1 & 0 & 0 \\
                      0 & -\case{r^2{ \gamma}_{\theta\theta}}{{2{ \gamma}_{rr}}} & 0 \\
		      0 & 0 & -\case{r^2{ \gamma}_{\theta\theta}\sin^2{\theta}}{{2{ \gamma}_{rr}}}
		      \end{array} \right) . 
\end{equation}
The vectorial quantities $\beta^a$, $\Lambda^a$, $\Delta\Gamma^a$ and $Z_a$ only have a radial non-zero component, that we will denote by $\beta^r$, $\Lambda^r$ and $Z_r$ for the variables and in the equations we will set 
\begin{equation}
\Delta\Gamma^i=\left( \case{2}{\gamma_{\theta\theta} r}- \case{2}{\gamma_{rr} r} + \case{\gamma_{rr}'}{2 \gamma_{rr}^2} -  \case{\gamma_{\theta\theta}'}{\gamma_{rr} \gamma_{\theta\theta}}, 0, 0 \right)^T  , 
\end{equation}
where we have used the component values of $\hat\gamma_{ab}$ given by $\hat\gamma_{ij}=diag(1,r^2,r^2\sin^2\theta)$. 
The full 4-dimensional line element is then 
\begin{equation}\label{nocomple}
\fl d\bar s^2 = - \left(\alpha^2-\chi^{-1}\gamma_{rr}{\beta^r}^2\right) dt^2 + \chi^{-1}\left(2\, \gamma_{rr}\beta^r dt\,dr +  \gamma_{rr}\, dr^2 +  \gamma_{\theta\theta}\, r^2\, d\sigma^2\right) ,
\end{equation}
where $d\sigma^2=d\theta^2+\sin^2\theta\,d\phi^2$ is the standard angular line element. 

The evolution equations for the scalar field in spherical symmetry are derived from the wave equation \eref{4dkgf}, converting it to a first order in time and second order in space hyperbolic system by introducing the auxiliary variable  $\Pi=\dot\Phi$. The complete spherically symmetric equations are contained in \ref{appspher}. Note that the equations presented are not simply the result of substituting the previous ansatz into the tensorial equations \eref{tensoreqs} and \eref{tensorceqs}, but the final equations, where the variable transformations and treatments that will be explained in section \ref{regularizingsec} have already been performed. 

\section{Initial data} \label{inihyp}

\subsection{Hyperboloidal foliations of Minkowski space}

We now construct a hyperboloidal version of our geometric setup in spherical symmetry \cite{Zenginoglu:2007jw}, starting with Minkowski spacetime. We define a hyperboloidal foliation of spacetime in terms of a new time coordinate $t$ that is related to the time coordinate on Cauchy surface, $\tilde t$, by means of a height function $h(\tilde r)$,
\begin{equation}\label{ttrafo}
t = \tilde t-h(\tilde r)  .
\end{equation}
The role of the height function is to ``raise'' the hypersurfaces $t=const.$ so that they reach $\scri^+$, and we use the simple one-parameter family  
\begin{equation}\label{hfunc}
h(\tilde r) = \sqrt{a^2+\tilde r^2}.
\end{equation}
For Minkowski spacetime, this height function results in a constant mean curvature foliation with $a=-3/{\tilde K_{0}}$, with $\tilde K_{0}$ the trace of the physical extrinsic curvature. Its value is a measure of how close the spacelike hyperboloidal slices come to light cones as they approach $\scri^+$: the smaller $\vert \tilde K_{0} \vert$ (the larger the value of $a$) the closer to a Cauchy slice we are, and the smaller the characteristic speeds of the system become (they are indirectly proportional to $a$, see table \ref{eigen}), so e.g. the time it takes waves to reach null infinity is larger. 
On the other hand, the larger $\vert \tilde K_{0} \vert$ (the smaller $a$), the more boosted the hyperboloidal slices are with respect to an observer at rest at spatial infinity and the larger the characteristic speeds. 
To include $\scri^+$ in the computational domain we compactify the radial coordinate $\tilde r$; we express it in terms of a new compactified $r$ and the time-independent conformal factor $\Omega=\Omega(r)$, 
\begin{equation}\label{rtrafo}
\tilde r=\frac{r}{\Omega}, \qquad \textrm{where} \quad r \rightarrow 0 \ \Leftrightarrow \ \tilde{r}\rightarrow 0 \quad \textrm{and} \quad r\rightarrow r_{\!\!\scri} \ \Leftrightarrow \ \tilde{r}\rightarrow\infty . 
\end{equation}
The constant $r_{\!\!\scri}$ denotes the fixed position of null infinity in our integration domain, which can be set to 1 without any loss of generality. The explicit form of $\Omega$ used here will be given in \eref{omegaexpr}. 

We will now express our initial data on the hyperboloidal foliation. We start with the standard Minkowski line element
\begin{equation}
d\tilde s^2 = -d\tilde t^2+{d\tilde r^2}+\tilde r^2 d\sigma^2 . 
\end{equation}
To obtain the same data on the hyperboloidal slice we transform the time and radial coordinates (in this order) using the relations \eref{ttrafo} and \eref{rtrafo}. Then we conformally rescale the line element $d\bar s^2=\Omega^2d\tilde s^2$: 
\begin{equation}\label{fsthyp}
\fl d\bar s^2= -\Omega^2 dt^2-2 \,h'\left(\Omega-r\,\Omega'\right)dt\,dr+\left(1- h'^2\right)\left(\frac{\Omega-r\,\Omega'}{\Omega}\right)^2d r^2 + r^2 d\sigma^2 .
\end{equation}
This is how our initial line element is expressed on the compactified hyperboloidal slice, where $h'$ is now interpreted as a function of $r/\Omega$, and $\Omega$ as function of the new rescaled $r$, as defined by \eref{rtrafo}. 

Comparing \eref{nocomple} with \eref{fsthyp} we assign the values for our metric components, choosing to maintain the radial coordinate as areal radius, that is $\gamma_{\theta\theta}=1$:
\begin{subequations} \label{varshypcomp}
\begin{eqnarray}
\chi &=& 1 , \\
\gamma_{rr} &=& \left(1-h'(r/\Omega)^2\right)\left(\frac{\Omega-r\,\Omega'}{\Omega}\right)^2 = \frac{a^2\left(\Omega-r\,\Omega'\right)^2}{r^2+a^2\Omega^2}  ,\label{grrini} \\
\gamma_{\theta\theta} &=& 1 , \\
\alpha &=& \sqrt{\frac{\Omega^2}{1-h'(r/\Omega)^2}} = \frac{\sqrt{r^2+a^2\Omega^2}}{a}  \label{inialpha}, \\
\beta^r &=& -\frac{\Omega^2 \,h'(r/\Omega)}{(1-h'(r/\Omega)^2)(\Omega-r\,\Omega')} = -\frac{r\sqrt{r^2+a^2\Omega^2}}{a^2\left(\Omega-r\,\Omega'\right)}   . \label{betarini}
\end{eqnarray}
\end{subequations}
For simplicity we use the following expression for the conformal factor, which is obtained by setting the last part of \eref{grrini} $=1$ and gives conformally flat initial data:
\begin{equation}\label{omegaexpr}
\Omega=\frac{r_{\!\!\scri}^2-r^2}{2\,a\, r_{\!\!\scri}}  .
\end{equation}
More information about conformal choices will be given in \sref{gaugecond}. 
The flat spacetime values for the rest of the evolution variables (derived from the previous ones) are 
\begin{equation}\label{varsderhypcomp}
A_{rr}=\Lambda^r=\Theta=0\ \quad\textrm{and}\quad K=-\frac{3}{a(\Omega-r\,\Omega')}  .
\end{equation}


\subsection{Gauge wave initial data}\label{initialgauge}

We set the values for flat spacetime as in \eref{varshypcomp} and \eref{varsderhypcomp} with the exception of $\alpha$, that takes the initial value in \eref{inialpha} plus a perturbation of the form
\begin{equation}\label{initialgaussian}
\delta\alpha_0=A_{\alpha}\e^{-\frac{(r^2-c^2)^2}{4\sigma^4}}  .
\end{equation}
Without scalar field the constraints are independent of the gauge variables, so that they are satisfied by this initial data. We choose this particular form for the initial perturbation because it is an even function at the origin (appropriate for a variable with even parity there, like $\alpha$ or $\Phi$). This is relevant if the perturbation is non-zero at the origin, although it is not the case in the simulations presented in this paper.  

\subsection{Scalar field initial data}\label{initialwave}
 
In this case all GBSSN/\CZ{} variables are set initially to their flat spacetime values excepting $\chi$, so that the metric will be conformally flat. The constraint $Z_r$ then vanishes and we set $\Pi_0=\beta_0^r\Phi'_0$ to satisfy the momentum constraint $\mathcal{M}_r$. 
Defining $\chi_0\equiv\psi^{-4}$ we get the standard form for the Hamiltonian constraint as a quasilinear elliptic equation for $\psi$:
\begin{equation} \label{inihamil}
\fl \mathcal{H} \frac{\psi^5}{8}=-\psi''-\psi'\left(\frac{2}{r}-\frac{\Omega'}{\Omega}\right)+\psi\left(\frac{\Omega'}{r\Omega}-\frac{3\Omega'^2}{4\Omega^2}+\frac{\Omega''}{2\Omega}\right)+\frac{3\psi^5}{4a^2\Omega^2}-\pi\psi(\Phi_0')^2=0  .
\end{equation}
This equation is solved for $\psi$ using a shooting-and-matching technique, for further details on the numerical methods see Sec. \ref{nummeth}. For $\Phi_0$ we use a Gaussian in $r^2$, 
\begin{equation}\label{initialgaussianphi}
\Phi_0=A_{\Phi}\e^{-\frac{(r^2-c^2)^2}{4\sigma^4}}  .
\end{equation}

\section{Gauge conditions}\label{gaugecond}

Looking at  (\ref{einsteinten}) and multiplying by $\Omega^2$,
one can see that for a vacuum spacetime, $G[\bar g]_{ab} = 0$,
$(\bar\nabla_c \Omega) \bar\nabla^c \Omega = 0$ at 
${\mycal I}$, which thus must consist of null surfaces.
The gauge freedom in the choice of the conformal
factor can be used to rescale the conformal factor $\Omega$ by some
$\omega > 0$ such that $\hat\Omega = \omega \Omega$, which introduces a new metric $\hat g_{ab}=\hat\Omega^2 \tilde g_{ab}=\omega^2\bar g_{ab}$.
One can use this conformal gauge freedom to achieve a preferred conformal gauge (see e.g.  Chpt. 11 of \cite{Wald}), where
\begin{equation}\label{eq:Bondi_conformal_gauge}
\hat\nabla_a \hat\nabla_b \hat\Omega = 0 \qquad \mbox{on ${\mycal I}^+$}.
\end{equation}
This conformal gauge implies, that the null tangent $\hat l^a = \hat g^{ab}\hat\nabla_b \hat \Omega$ to the null geodesic generators of
${\mycal I}$ satisfies the affinely parameterized geodesic equation,
\begin{equation}\label{eq:geodesic_generators}
\hat l^a \hat\nabla_a \hat l^b = 0 \; .
\end{equation}
Consequently, expansion of the generators of ${\mycal I}$ vanishes in
addition to the shear and twist ($\hat l_a$ is a gradient).
The affine parameter $t_B$  of the null geodesic generators, scaled
such that $(\partial/\partial t_B)^a \hat\nabla_a t_B = 1$  is
generally known as Bondi parameter or Bondi time. The preferred conformal gauge also ensures that the three conformal factor terms in \eref{einsteinten} are regular at null infinity.

Following the approach of Zengino\u{g}lu \cite{Zenginoglu:2007it,Zenginoglu:2008pw}, we use a fixed conformal factor $\Omega$. It is coupled to the coordinate gauge by being explicitly expressed in terms of the radial coordinate in \eref{omegaexpr} and has the effect of fixing the coordinate location of null infinity and translating its geometrical properties into coordinate conditions and gauge conditions. In our present work the preferred conformal gauge condition \eref{eq:Bondi_conformal_gauge} will however only be satisfied by the initial data, and not at all times during evolution. As discussed in section \ref{resu}, we are able to construct a stable and convergent numerical scheme even in the absence of the preferred conformal gauge, and leave using this further regularization to future work.

\subsection{Reparametrization of null infinity by Bondi time} \label{bondi}

In order to parametrize null infinity by Bondi time in the current setup (adapting to the preferred conformal gauge), we recalculate the time variation of the conformal factor after the numerical simulation and rescale the time coordinate to obtain the Bondi time at $\scri^+$. The relation between the time-dependent rescaling $\omega$ of the conformal factor and the evolution variables at $\scri^+$ is 
\begin{equation}
\frac{\dot\omega}{\omega}=\frac{\beta^r\gamma_{rr}'}{2\gamma_{rr}}-\frac{\alpha\chi\alpha'}{\gamma_{rr}\beta^r}+{\beta^r}'-\frac{\beta^r\chi'}{2\chi}  .
\end{equation}
In the preferred conformal gauge the affinely parametrized time ($t_B$) is obtained by the coordinate change $dt_B = \frac{\hat\alpha^2}{\hat\beta^r\hat\Omega'} dt$, where $t$ is the coordinate time of the simulation and the gauge variables in the preferred conformal picture transform as $\hat\alpha=\omega\alpha$ and $\hat\beta^r=\beta^r$.
After the substitution $\hat\Omega'|_{\scri^+}=\omega\Omega'$, the Bondi time $t_B$ is obtained from
\begin{equation}
dt_B = \frac{\alpha^2\omega}{\beta^r\Omega'} dt .
\end{equation}
A comparison of the signal at code time $t$ and Bondi time $t_B$ is presented in figure \ref{phiscri}.

\subsection{Scri-fixing}

The scri-fixing condition consists of fixing the location of $\scri^+$ to a certain position in the integration domain. To do so in spherical symmetry, we choose the coordinate time vector to flow along $\scri^+$, that is, $\left(\case{\partial}{\partial t}\right)^a=\alpha \bar n^a+\beta^a$ is chosen to be an ingoing null vector at $\scri^+$.  In a more general geometric setup, we could also have chosen a shift vector which has non-radial components, including at null infinity.
\begin{figure}[htbp]
\center
	\begin{tikzpicture}[scale=1.2]
		\draw (0cm, 0cm) -- (8cm, 0cm); \draw [thick,->] (5cm, 0cm) -- (1.5cm, 3.5cm);
		\draw (1.3cm, 3.9cm) node {$\Omega=0$}; \draw (3.3cm, 2.4cm) node {$\scri^+$};
		\draw [very thick,->] (1cm, 0cm) -- (2cm, 0cm); \draw [very thick,->] (1cm, 0cm) -- (1cm, 1cm);
		\draw (0.7cm, 0.6cm) node {$\bar n^a$}; \draw (1.6cm, 0.3cm) node {$\bar r^a$};
		\draw [very thick,->] (5cm, 0cm) -- (4cm, 1cm); \draw [very thick,->] (5cm, 0cm) -- (6cm, 1cm);
		\draw (4.3cm, 1.3cm) node {$l^a$}; \draw (6.3cm, 1.3cm) node {$k^a$}; 
		\draw [very thick,->] (6.5cm, 2.8cm) -- (5.5cm, 3.8cm); \draw [very thick,->] (6.5cm, 2.8cm) -- (6.5cm, 3.8cm); \draw [very thick,->] (6.5cm, 2.8cm) -- (5.5cm, 2.8cm);
		\draw (4.9cm, 3.8cm) node {$\left(\frac{\partial}{\partial t}\right)^a$}; \draw (7cm, 3.3cm) node {$\alpha\bar n^a$}; \draw (6cm, 2.4cm) node {$\beta^a$};
	\end{tikzpicture}
	\caption{Timelike, spacelike and null vectors near $\scri^+$.}\label{scridiagram}
\end{figure}
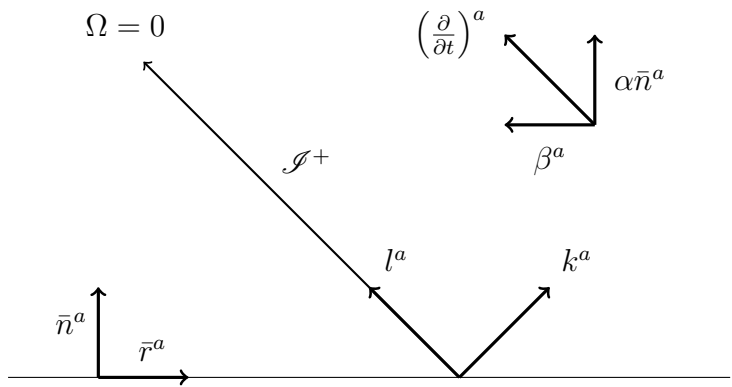

In our formulation $\scri^+$ is given by $\Omega=0$. Calling $l^a$ the null vector along $\scri^+$ (see \fref{scridiagram} for a graphical representation), the condition for scri-fixing $\left.\left(\frac{\partial}{\partial t}\right)^a\right|_{\scri}\equiv l^a$ is 
\begin{equation}
\left.\left(\frac{\partial}{\partial t}\right)^a\left(\frac{\partial}{\partial t}\right)_a\right|_{\scri}=
\left.\left(\nabla^a\Omega\right)\left(\nabla_a\Omega\right)\right|_{\scri}=0. 
\end{equation}
Since $\scri^+$ is a null surface, $l^a$ is parallel to $\nabla^a\Omega$, and we will choose our shift vector such that they are also parallel to $\left(\frac{\partial}{\partial t}\right)^a$ at $\scri^+$. 
Consequently, we obtain $\left.\partial_t\Omega\right|_{\scri}=0$, consistent with the value of $\Omega$ not being evolved in time. 

We will choose our shift vector on $\scri^+$ according to 
\begin{equation}
 \left.-\alpha^2+\beta^a\beta_a\right|_{\scri}=0  , \quad \textrm{in spherical symmetry:}  \left.-\alpha^2+\chi^{-1}\gamma_{rr}{\beta^r}^2\right|_{\scri}=0  .
\end{equation}
Comparing with \eref{fsthyp} and \eref{nocomple} we can see that the flat spacetime values satisfy the previous condition. For simplicity we set $\beta^r$ to its flat spacetime value \eref{betarini}, which has the appropriate value at future null infinity to keep $\scri^+$ fixed in its place, and maintain this value throughout the evolution without evolving the shift. 
The scri-fixing condition thus takes the form 
\begin{equation}
\beta^r=-\frac{r\sqrt{r^2+a^2\Omega^2}}{a^2\left(\Omega-r\,\Omega'\right)}=-\frac{r}{a}  ,
\end{equation}
where we have substituted \eref{omegaexpr} to obtain the simpler expression. 
This initial value of $\beta^r$ satisfies both the scri-fixing and the constant areal radius conditions.

\subsection{Generalized Bona-Massó family of slicing conditions}

Evolving on hyperboloidal foliations requires extra freedom in how the slicing condition is imposed. We generalize the Bona-Massó slicing condition \cite{Bona:1994dr} to include a source term $L_0$ in addition to the free functions $f(\alpha)$ and $K_0$:
\begin{equation}\label{lapseq}
\dot \alpha=\beta^r\alpha'-f(\alpha)\left(\bar K-K_0\right)+L_0  .
\end{equation}
In the following we will restrict ourselves to a generalized harmonic slicing, $f(\alpha)=\alpha^2$, which avoids unnecessary unphysical propagation speeds (in particular at null infinity), and has been demonstrated to be appropriate for hyperboloidal evolution in \cite{Zenginoglu:2007it}. The ``1+log'' case popular in astrophysical black hole evolutions may more easily lead to some obstructions for hyperboloidal slices as discussed in \cite{Ohme:2009gn}.

The presence of $K_0$ in the lapse's equation of motion is necessary to prevent the lapse from growing exponentially, as the flat spacetime value of the extrinsic curvature $K$ on the hyperboloidal foliation is negative. Thus $K_0$ has to be set analytically in such a way that the coefficient in front of $\alpha^2$ is negative. We will later replace $\bar K$ by a new evolution variable $\K$  defined below in \eref{DeltaKdef} to suppress continuum instabilities, and we will choose $K_0$ in such a way that the coefficient in front of $\alpha^2$ becomes $-({\K}/{\Omega}+\xi_{\alpha})$, where $\xi_{\alpha}$ is a constant parameter. This gives $K_0 = -\xi_{\alpha}-{3}/(a \Omega)$ in the final equation. The value of $\xi_\alpha$ is set to achieve stability, we have  found $\xi_\alpha\in[1,4]$ to be a good choice. For a more complete approach to analyze continuum instabilities, which was not necessary in our case, see \cite{Frauendiener:2004bj}.
Once $K_0$ is set, the source function $L_0$ is calculated from flat spacetime initial data on the hyperboloidal foliation. In our case we obtain: 
\begin{equation}
 L_0= \xi_{\alpha}\left(\frac{\sqrt{r^2+a^2\Omega^2}}{a}\right)^2  + \frac{2 {\beta^r} \sqrt{r^2+ a ^2 \Omega ^2} \Omega '}{ a  \Omega }  -\frac{r{\beta^r} }{ a^2\Omega}  .
\end{equation}

\section{Hyperbolicity analysis} \label{hypanal}

The GBSSN and \CZ{} systems of evolution equations are systems of quasilinear partial differential equations.
In order to understand the well-posedness and propagation properties of their associated initial value problems, we perform a hyperbolicity analysis of the principal parts of the GBSSN and \CZ{} systems.

A first order in time and second order in space system like the ones we consider here can be analyzed following 
\cite{Sarbach:2002bt,Nagy:2004td,Gundlach:2004ri,Gundlach:2004jp,Beyer:2004sv,Calabrese:2005ft},
where the evolution variables are classified into those which are only differentiated once in space and those which are differentiated twice. Here the metric variables and the scalar field belong to the second group and the rest of variables, to the first one. The fields that appear in the evolution equations are $\chi, \gamma_{rr}, \gamma_{\theta\theta}, \alpha, \Phi, A_{rr}, K, \Lambda^r, \Theta \textrm{ and }  \Pi$. The characteristic fields (also called eigenfields), which are defined through a first order reduction, additionally contain the spatial derivatives $\chi', \gamma_{rr}', \gamma_{\theta\theta}', \alpha' \textrm{ and } \Phi'$. Note that the value of $\beta^r$ is fixed during evolution in the present work.
The analysis of the principal part follows \cite{Calabrese:2005ft} and the matrix that includes the principal part information is given by equation (34) there. 

A complete set of eigenvectors is found and all characteristic speeds (also called eigenspeeds) are real, thus the continuum system is strongly hyperbolic and, as we are working in spherical symmetry, it is automatically symmetric hyperbolic. For generality we have calculated the characteristic fields and characteristic speeds of the equations keeping the metric component $\gamma_{\theta\theta}$ as an evolution variable, but we can eliminate it using the freedom of the determinant of the conformal metric. 
Indeed it is after fixing this degree of freedom that we obtain the standard Z4c/CCZ4 formulation.
In this case, the first characteristic field of \tref{eigen} vanishes and in the rest $\gamma_{\theta\theta}$ is substituted by $\gamma_{\theta\theta}=1/{\sqrt{\gamma_{rr}}}$.
This is in agreement with the initial data choices in \eref{varshypcomp} after substitution of \eref{omegaexpr}. The determinant of the spatial metric after elimination of $\gamma_{\theta\theta}$ is $\chi^{-3}$. 

\lineup
\begin{table}[hhh]
\caption{\label{eigen}Characteristic fields and speeds, using $c_0=-\beta^r$ and $c_\pm=-\beta^r\pm\alpha\sqrt{\case{\chi}{\gamma_{rr}}}$ to simplify the notation. ``GBSSN / \CZ{}'' denotes the eigenquantities common to both formulations,  whereas  ``GBSSN'' and ``\CZ{}'' correspond only to the mentioned system.}
\begin{tabular}{@{}llcc}
\br
System&Eigenfields&Eigenspeeds&at $\scri^+$\\
\mr
GBSSN / \CZ{}&$\frac{\gamma_{rr}'}{\gamma_{rr}}+\frac{2\gamma_{\theta\theta}'}{\gamma_{\theta\theta}}$&0&0\\
GBSSN / \CZ{}&$-2\gamma_{rr}\Lambda^r+\frac{2\alpha'}{\alpha}-\frac{\chi'}{\chi}$&$c_0$&$\frac{r_{\!\!\scri}}{a}$\\
GBSSN / \CZ{}&$2\gamma_{rr}\Lambda^r+\frac{3}{\sqrt{\gamma_{rr}\chi}}A_{rr}-\frac{\gamma_{rr}'}{\gamma_{rr}}+\frac{\gamma_{\theta\theta}'}{\gamma_{\theta\theta}}-\frac{2\alpha'}{\alpha}+\frac{\chi'}{\chi}$&$c_-$&0\\
GBSSN / \CZ{}&$2\gamma_{rr}\Lambda^r-\frac{3}{\sqrt{\gamma_{rr}\chi}}A_{rr}-\frac{\gamma_{rr}'}{\gamma_{rr}}+\frac{\gamma_{\theta\theta}'}{\gamma_{\theta\theta}}-\frac{2\alpha'}{\alpha}+\frac{\chi'}{\chi}$&$c_+$&$\frac{2\,r_{\!\!\scri}}{a}$\\ 
GBSSN / \CZ{}&$\frac{\alpha'}{\alpha}-\sqrt{\frac{\gamma_{rr}}{\chi}}K$&$c_-$&0\\
GBSSN / \CZ{}&$\frac{\alpha'}{\alpha}+\sqrt{\frac{\gamma_{rr}}{\chi}}K$&$c_+$&$\frac{2\,r_{\!\!\scri}}{a}$\\
GBSSN&$\frac{3\chi'}{\chi}+\frac{2\alpha'}{\alpha}-\frac{\gamma_{rr}'}{\gamma_{rr}}-\frac{2\gamma_{\theta\theta}'}{\gamma_{\theta\theta}}$&$c_0$&$\frac{r_{\!\!\scri}}{a}$\\
\CZ{}&$\gamma_{rr}\Lambda^r-\frac{\gamma_{rr}'}{2\gamma_{rr}}-\frac{\gamma_{\theta\theta}'}{\gamma_{\theta\theta}}+\frac{2\chi'}{\chi}+2\sqrt{\frac{\gamma_{rr}}{\chi}}\Theta$&$c_-$&0\\
\CZ{}&$\gamma_{rr}\Lambda^r-\frac{\gamma_{rr}'}{2\gamma_{rr}}-\frac{\gamma_{\theta\theta}'}{\gamma_{\theta\theta}}+\frac{2\chi'}{\chi}-2\sqrt{\frac{\gamma_{rr}}{\chi}}\Theta$&$c_+$&$\frac{2\,r_{\!\!\scri}}{a}$\\
Scalar field&$\Phi'+\frac{1}{\alpha}\sqrt{\frac{\gamma_{rr}}{\chi}}(\Pi-\beta^r\Phi')$&$c_-$&0\\
Scalar field&$\Phi'-\frac{1}{\alpha}\sqrt{\frac{\gamma_{rr}}{\chi}}(\Pi-\beta^r\Phi')$&$c_+$&$\frac{2\,r_{\!\!\scri}}{a}$\\
\br
\end{tabular}
\end{table}

\begin{figure}[htbp!!]
\center
\mbox{\includegraphics[width=0.65\linewidth]{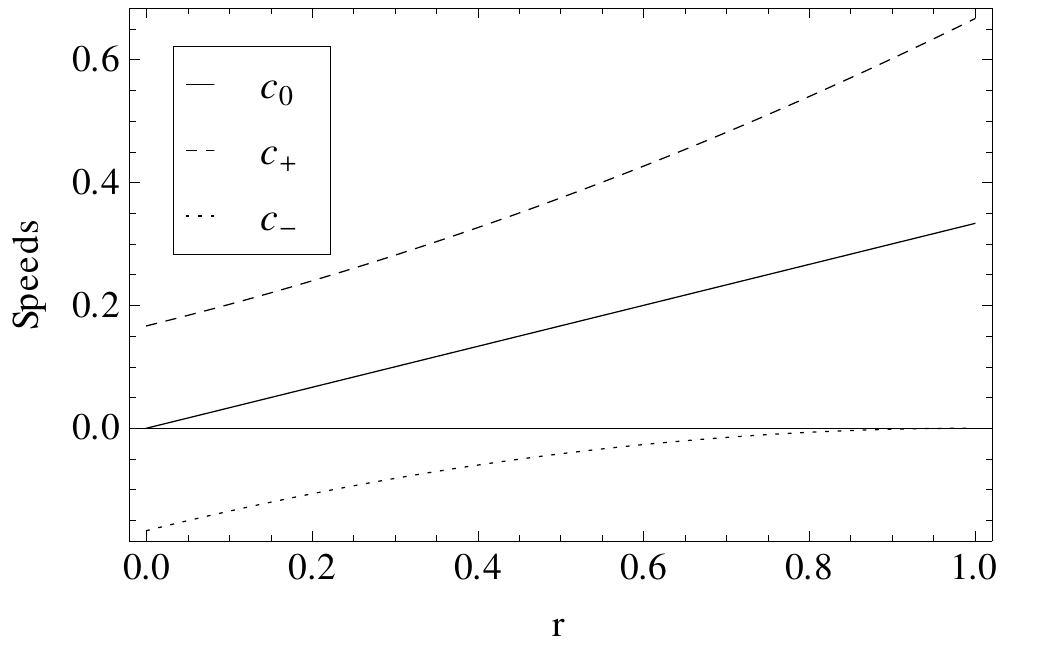}}\vspace{-1.5ex}
\caption{The zero speed $c_0$ is $-\beta^r$ and the lightspeeds are $c_\pm=-\beta^r\pm\alpha\sqrt{\case{\chi}{\gamma_{rr}}}$. The flat spacetime values of the variables are used, $\Omega$ is set as in \eref{omegaexpr} and $r_{\!\!\scri}=1$ and $a=1$.}
\label{lightspeeds}
\end{figure}

As expected and shown in figure \ref{lightspeeds} and in the last column of \tref{eigen}, the characteristic speeds at $\scri^+$ are either zero or positive, so that no incoming modes exist.  A complete list of the characteristic speeds and fields is presented in \tref{eigen}, to which the characteristic fields $\chi$, $\gamma_{rr}$, $\gamma_{\theta\theta}$, $\alpha$ and $\Phi$ have to be added with zero propagation speed. The principal part of the scalar field equations decouple from the rest of the system. This eigendecomposition can be compared with those in \cite{Brown:2007nt,Field:2010mn}, substituting our $\Lambda^r$ by $\Gamma^r$, rearranging some of the eigenvectors, omitting the scalar field part and taking into account that in the references the 1+log slicing condition was used instead of the harmonic one and the shift was part of the evolution system.

All characteristic speeds have one of four values: zero,  equal in magnitude to the shift ($c_0$), or equal to the speed of light for incoming ($c_-$, $c_-\vert_{\scri}=0$) or outgoing ($c_+$) radial null rays. The three speeds $(c_0, c_-, c_+)$ are plotted as functions of the compactified radial coordinate in figure \ref{lightspeeds}, and are
 inversely proportional to the parameter $a$.

\section{Numerical methods} \label{nummeth}

The spherically symmetric equations are implemented in a 1-dimensional numerical code that uses the Method of Lines approach. A standard 4th order Runge-Kutta scheme is used for the time integration. The spatial derivatives are discretized with standard finite differences of orders from 2 to 8, but for the simulations presented here we use 4th order finite differences. 

The initial data for the gauge waves is simply set analytically in the numerical variables, but in the scalar field case the Hamiltonian constraint \eref{inihamil} is solved numerically using the shooting-and-matching technique, integrating with a 4th order Runge-Kutta scheme until the condition of asymptotic flatness is met.
The integration domain extends from the origin ($r=0$) to future null infinity, which in our case is placed at $r=r_{\!\!\scri}$. 

Some terms in the equations diverge at $r=0$ and $r=r_{\!\!\scri}$. In the present work we use a staggered grid, where the numerical grid is moved half a spatial step to the right and the outermost point is eliminated, so that the two divergent points (the origin and null infinity) are avoided. No further treatment is required for the initial data or the equations. In order to compute the solution exactly on $\scri^+$ we extrapolate half a grid point using extrapolation of an order at least as high as the convergence order used in the simulation to obtain the signal at the desired place and we see that it converges properly (see figure \ref{phiscri}). The stable implementation of a non-staggered grid, where the divergent terms in the RHSs are evaluated using the l'H\^opital rule to obtain regular expressions at $r=0$ and $r=r_{\!\!\scri}$, is current work. 

The ghost points are filled in a different way at the origin and at the outer boundary:
\begin{itemize}
\item $r=0$: the ghost points ``to the left of the origin'' are filled according to the parity of the corresponding variable. For even variables ($\chi$, $\gamma_{rr}$, $\gamma_{\theta\theta}$, $A_{rr}$, $K$, $\alpha$, $\Phi$, $\Pi$ and $\Theta$) the value of the ghost points is the reflection of the values of the points on the right of the origin, whereas the odd variables ($\Lambda^r$ and $\beta^r$) take minus the value of their reflected point. 
\item $r=r_{\!\!\scri}$: the ghost points beyond $\scri^+$ are filled using the outflow boundary conditions in \cite{Calabrese:2005fp}. We observe that it is advantageous to extrapolate with a higher order to the one corresponding to the finite differences. 
\end{itemize}
In the derivatives evaluated next to the boundaries we can use centered stencils, taking advantage of the ghost points, or use one-sided stencils, so that the ghost points are not needed. No relevant differences were observed between the two methods, so we chose the centered stencils at $\scri^+$.

As expected for nonlinear equations, the discrete evolution equations require some dissipation in order to remain numerically stable. For this reason Kreiss-Oliger dissipation terms \cite{kreiss1973methods} as used in \cite{Babiuc:2007vr} with operator $Q$ of order $2n$ (suitable for a $2n-2$ convergence finite difference setup), 
\begin{equation}
Q=\epsilon\,(-1)^n(\Delta r)^{2n-1}\frac{D_+^nD_-^n}{2^{2n}} \quad \textrm{with} \quad D_\pm A_i = \pm\frac{A_{i\pm1}-A_i}{\Delta r}  ,
\end{equation}
are added to the RHSs of the equations of motion, $\partial_tA\to\partial_tA+QA$. To avoid affecting the convergence order of the spatial derivatives, the order of convergence used in the dissipation terms is 2 orders higher. This requires an extra ghost point. The Kreiss-Oliger dissipation that we use is defined using a centered stencil, so that it needs the information from the ghost points, which is obtained by extrapolation, as mentioned above. 
Another option is not to set any dissipation on the outermost points of the integration domain to avoid using the boundary conditions. 
Some configurations of evolution equations and parameters are quite unaffected by the dissipation choice we make at the boundary, but it can have larger effects on other configurations. 
The larger the initial perturbation is, the more dissipation is needed to keep it stable. Nevertheless, the less dissipation used, the better the convergence results will be.

It is customary in numerical relativity to use off-centered stencils for advection-type derivatives arising from the shift vector, see e.g.~\cite{Husa:2007hp}, where this has been found essential for the accuracy of  ``moving puncture'' binary black hole evolutions with sixth order accurate stencils, and the discussion regarding the shifted wave equation in 
 \cite{Chirvasa:2008xx}. We have therefore also implemented these special one-point forward off-centered stencils in the derivatives included in the advection terms of the equations. The presence of these off-centered stencils in the advection terms decreases the amount of dissipation needed to keep the simulation stable and can increase accuracy, however we found a tradeoff with how clean convergence tests look. We will investigate this in the future in the context of black hole evolutions.

\section{Stability analysis}\label{regularizingsec}

A necessary condition to give meaning to the numerical solution of a time evolution problem, is that this problem is well posed in both its continuum and discrete formulations, where discrete well-posedness is usually referred to as numerical stability, see e.g. \cite{GusKreOlig,lrr-1998-3,Calabrese:2005ft}. In our formulation of the hyperboloidal initial value problem, where the equations become singular at null infinity, we do not have a rigorous proof of continuum well-posedness or numerical stability, and ultimately rely on convergence tests. However, well-posedness is not sufficient to guarantee that a given problem can be practically solved with numerical methods. The Einstein equations, in particular, are prone to exhibit unphysical continuum instabilities, as is the case for other gauge theories, since our evolution variables contain information about the gauge and constraint violation. Even if a solution of the equations is ``well-behaved'' regarding the physical degrees of freedom, instabilities can easily arise in the unphysical degrees of freedom. Here, with ``well-behaved'' we describe the situation that we can produce accurate long-term numerical evolutions which show the expected numerical convergence behaviour.
Unsurprisingly, a direct implementation of the spherically symmetric equations results in simulations which are not well-behaved. In order to track down and cure the instabilities we followed the steps discussed below.

\subsection{Analysis by subsystems}


Dividing the original system of equations into simpler hyperbolic subsystems enables us to find problems in a hierarchical way. 
The first check is performed evolving the single equations: for each of the equations of motion we substitute the initial values of all variables excepting the one which is evolved, so that all equations decouple. 
For a variable $A$ that translates to evolving 
\begin{equation}\dot A = f(A,A',r,\textrm{parameters})  .
\end{equation}
Note that for these mixed order hyperbolic systems, second spatial derivatives of a variable do not appear in the RHS of that variable's equation of motion.
This is especially useful for detecting exponential growths, that appear in the equation's RHS in the form of terms of the form $\lambda A$ with $\lambda>0$. 

We eliminate the $\gamma_{\theta\theta}$ variable by substituting it by $\gamma_{\theta\theta}\equiv \gamma_{rr}^{-1/2}$. This takes advantage of the degree of freedom given by the determinant, setting it to unity, and also uses the fact that the metric initial data are conformally flat. We mostly concentrate on the GBSSN system of equations, as it uses one variable less than the \CZ{} one ($\Theta$), and we evolve the vacuum solution (no scalar field coupled). 

The subsystem $(\chi , \gamma_{rr} , A_{rr} , \Lambda^r)$ is the largest hyperbolic subsystem that can be constructed with these variables and is well-behaved in the above sense, i.e. the numerical solution converges at the appropriate order and no continuum instabilities can be detected, thus small initial perturbations with respect to the stationary state disappear during the evolution and the subsystem arrives at its stationary state.

We detect an exponential growth in the subsystem $(\alpha , K)$, due to a term of the form $\case{\lambda}{\Omega} K$ with $\lambda>0$ in $\dot K$'s RHS. This can be solved with the variable transformation for $K$ described in the next subsection. Once the exponential growth has disappeared, the subsystem $(\chi,\alpha,\K)$ is also well-behaved. 

\subsection{Variable transformations} \label{vartrans}

\subsubsection{Transformation for $K$:}

The variable transformation that we perform on $K$ and that solves the stability problem is based on transforming back to the trace of the physical extrinsic curvature. The relation between the traces of the physical $\tilde K_{ab}$ and conformal $\bar K_{ab}$ extrinsic curvature is
\begin{equation}
\bar K = \frac{1}{\Omega}\left(\tilde K-\frac{3\partial_\perp\Omega}{\alpha}\right)=\frac{1}{\Omega}\left(\tilde K+\frac{3\,\beta^a\partial_a\Omega}{\alpha}\right)  .
\end{equation}
We then define a new evolution variable $\K$ as
\begin{equation}\label{DeltaKdef}
\K=\Omega K-\tilde K_0-\frac{3\,\beta^r\Omega'}{\alpha}.
\end{equation}
Here we set $\tilde K_0=-\case{3}{a}$, the flat spacetime value of the trace of the physical extrinsic curvature, such that now the Minkowski value of the variable $\K$ is simply zero. Note that this change also makes the constraints independent of lapse and shift in the vacuum equations for our choice of variables. 

Although not required to solve any stability issues, for consistency with the transformation of the trace of the extrinsic curvature we rescale the Z4 variable $\Theta$ to a new quantity $\Te$ according to
\begin{equation}
\Te=\Omega\Theta .
\end{equation}
Note that the characteristic fields presented in table \ref{eigen} are expressed in terms of $K$ and $\Theta$. To obtain them in terms of the new variables, substitute $K=\K/\Omega$ and $\Theta=\Te/\Omega$ and multiply the complete characteristic field by $\Omega$.

\subsubsection{Variable choice for the scalar field:} \label{phichoice}

The amplitude of the scalar field $\Phi$ and its time derivative $\Pi$ becomes zero at null infinity (it is inversely proportional to the distance). We can use as evolution variables either $\case{r}{\Omega}\Phi$ or $\case{\Phi}{\Omega}$ (and the same for $\Pi$): the first option changes the parity of the two variables at the origin and also forces them to be zero there, so that we chose the second option: our evolution variables will be $\bar\Phi={\Phi}/{\Omega}$ and $\bar\Pi={\Pi}/{\Omega}$. 
We implemented the scalar field equations using two different definitions for the auxiliary variable of the rescaled scalar field, $\bar\Pi\equiv\dot{\bar\Phi}$ and $\bar\Pi\equiv\case{1}{\alpha}\left(\dot{\bar\Phi}-\beta^r\bar{\Phi}'\right)$. Here we chose to show the simulations using $\bar\Pi\equiv\dot{\bar\Phi}$, because its performance for the convergence test was slightly better.

\subsubsection{Other transformations:}

Variable transformations that simplify the initial and stationary values of the gauge variables (convenient for visualization purposes, but which do not affect the stability of the code) are: 
\begin{equation}\label{albettrans}
\alpha_{unity}=\alpha\frac{a}{\sqrt{r^2+a^2\Omega^2}}\qquad \textrm{and}\qquad \beta^r_{zero}=\beta^r+\frac{r\sqrt{r^2+a^2\Omega^2}}{a^2\left(\Omega-r\,\Omega'\right)}  , 
\end{equation}
where we transform lapse and shift with the help of their values for flat spacetime \eref{varshypcomp}. Using this the initial value of $\alpha_{unity}$ equals one and the fixed value of $\beta^r_{zero}$ is zero.

\subsubsection*{}

Even after performing these variable transformations, the complete system is not yet well-behaved. What solves the problem is the addition of some specific damping terms, as we describe in the next subsection.  

\subsection{Constraint damping}\label{constraintdamping}

Many of the final equations possess a damping term of the form $\dot X=\frac{\lambda}{\Omega}X$ in their RHSs, with $X$ some evolution variable and $\lambda$ a coefficient that depends on other evolution variables and on $r$. In \tref{damp} we show the value of these $\lambda$s for the GBSSN equations substituting the flat spacetime values of the variables and evaluating them at $\scri^+$ ($r=r_{\!\!\scri}$). A heuristic stability requirement is that these coefficients, that accompany the terms that diverge at $\scri^+$, have to be negative everywhere ($\Omega$ is always positive in the physical domain).
\begin{table}[hhh]
\center
\caption{\label{damp} Values at $\scri^+$ of the coefficients over $\Omega$ that precede the variables.}
\begin{tabular}{@{}ccccccccc}
\br
$\dot \chi$&$\dot \gamma_{rr}$&$\dot \gamma_{\theta\theta}$&$\dot A_{rr}$&$\dot{\K}$&$\dot \Lambda^r$&$\dot \alpha$&$\dot \Phi$&$\dot \Pi$\\
\mr
0&0&0&$-\frac{2 r_{\!\!\scri}}{a^2}$&$-\frac{2 r_{\!\!\scri}}{a^2}$&0&$-\frac{3 r_{\!\!\scri}}{a^2}$&0&$-\frac{2 r_{\!\!\scri}}{a^2}$\\
\br
\end{tabular}
\end{table}

The evolution equation of $\Lambda^r$ does not have a damping term like $A_{rr}$, $\K$ and $\Pi$, and indeed we find that we need to add a damping term with the appropriate coefficient to achieve stability. This can be easily done by keeping the Z4 damping term in $\dot \Lambda^r$ even when we use the GBSSN equations: 
\begin{equation}\label{lambdarterm}
\fl\dot \Lambda^r=\textrm{GBSSN-RHS} - \frac{2\kappa_1 \alpha Z_r}{\gamma_{rr}\Omega}, \quad \textrm{being} \  Z_{r}=\frac{1}{2}\gamma_{rr}\Lambda^r+\frac{1}{r}-\frac{\gamma_{rr}}{\gamma_{\theta\theta}r}-\frac{\gamma_{rr}'}{4\gamma_{rr}}+\frac{\gamma_{\theta\theta}'}{2\gamma_{\theta\theta}}  .
\end{equation}
This would be called ``Z3 damping'' in \cite{Gundlach:2005eh}.
The complete \CZ{} system is well-behaved, and indeed an equivalent damping term with the appropriate sign also appears in $\dot\Theta$'s RHS. 

We note that if we would extend the computational domain beyond future null infinity, we would face the problem that the conformal factor terms change sign for $r>r_{\!\!\scri}$ ($\Omega(r>r_{\!\!\scri})<0$) and the damping terms that we have just discussed are likely to cause exponential growth instead of damping. Instabilities in the continuum equations beyond null infinity may have to be treated with methods that go beyond what is presented here.

\subsection{Regularity conditions}\label{regularitysubsec}

The regularity conditions are the relations between the evolution variables that satisfy that the equations are regular at $\scri^+$. Although the conformal factor terms are analytically regular, this does not always carry over to the discretized equations and thus the conditions may have to be reinforced (this is not done in our experiments). To find them we simply look at the numerator of the terms divided by the conformal factor $\Omega$ that appear in the RHSs. The regularity conditions for the GBSSN system are: 
\begin{itemize}
\item $\left.-\alpha^2+\chi^{-1}\gamma_{rr}{\beta^r}^2\ \right|_{\scri}=0$. This is the scri-fixing condition, the one that makes the time coordinate null at $\scri^+$. 

‪\item $\left.\K-\frac{3}{a} \ \right|_{\scri}=\left.-\frac{3\beta^r\Omega'}{\alpha}\ \right|_{\scri}=\left.+\frac{3\sqrt{\chi}\,\Omega'}{\sqrt{\gamma_{rr}}}\ \right|_{\scri}$, obtained from $\dot \chi$ and $\dot \K$ due to the transformation of the trace of the extrinsic curvature. In the second equality we have used the scri-fixing condition. 

\item $\left.\K' \right|_{\scri}=\left.-\frac{3A_{rr}\Omega'}{\gamma_{rr}} \ \right|_{\scri}$, from $\dot \Lambda^r$. 

\item $\left.A_{rr}\right|_{\scri}=\left.-\frac{\alpha\chi}{3\beta^r}\left( \frac{2}{r}+\frac{\gamma_{rr}'}{\gamma_{rr}}+\frac{\gamma_{\theta\theta}'}{\gamma_{\theta\theta}}-\frac{2\chi'}{\chi}-\frac{2\Omega''}{\Omega'}\right) \ \right|_{\scri}$, calculated from $\dot A_{rr}$. 

\item There is an extra condition that arises from the $\K$ condition above, the fixed value of $\beta^r$ and the harmonic slicing condition implemented, and that puts an extra constraint on the ones already listed: $\left.\alpha\ \right|_{\scri}=\frac{r_{\!\!\scri}}{a}$, obtained by substituting the value of the conformal factor \eref{omegaexpr}. Setting the values of $\alpha$, $\beta^r$ and the conformal factor at $\scri^+$, the previous conditions turn into: 
\begin{itemize}
\item $\left.\chi \right|_{\scri}=\left.\gamma_{rr} \right|_{\scri}$ ,
\item $\left.\K \right|_{\scri}=0$ ,
\item $\left.\K' \right|_{\scri}=\left.-\frac{3A_{rr}\Omega'}{\gamma_{rr}} \ \right|_{\scri}$ unchanged,
\item $\left.A_{rr}\right|_{\scri}=\left.\frac{1}{3}\left( \gamma_{rr}'+\frac{\gamma_{rr}\gamma_{\theta\theta}'}{\gamma_{\theta\theta}}-2\chi'\right) \ \right|_{\scri}$ .
\end{itemize}

\item $\left.\Pi \right|_{\scri}=0$ for the auxiliary variable of the scalar field. However, the rescaled auxiliary variable, $\bar\Pi=\Pi/\Omega$, is allowed to take non-zero values at null infinity. 

\end{itemize}

\section{Results}\label{resu}

The main result is that we have achieved a numerically stable evolution of the hyperboloidal initial value problem in spherical symmetry.

\subsection{Evolution setup}

The system of equations evolved is the one presented in \ref{appspher}, but transforming the gauge variable $\alpha$ as in \eref{albettrans} and choosing a fixed shift, where $\beta^r=-r/a$ is set to its flat spacetime value. For convenience and as explained in subsection \ref{phichoice}, we also evolve the rescaled scalar fields $\Phi/\Omega$ and $\Pi/\Omega$, because their amplitudes are non-zero at future null infinity.

In the simulations shown here we used 4th order finite differences and the values $r_{\!\!\scri}=1$ and $a=1$. For the damping parameters we set $\xi_\alpha=1$, $\kappa_1=1.5$ and $\kappa_2=0$. The value $\kappa_1=1.5$ used here was appropriate for both GBSSN and \CZ{} formulations in our simulations, but choosing a smaller value $\kappa_1\sim1$ seemed to be preferred by \CZ{} in some cases. This value is bigger than the one suggested by \cite{Weyhausen:2011cg}, $\kappa_1\sim0.02$ also in spherical symmetry, but relatively close to the one used by \cite{Alic:2011gg}, $\kappa_1\sim1$.
The Kreiss-Oliger dissipation parameter is taken as $\epsilon=0.5$. 
The time-step was $\Delta t=5\cdot10^{-4}$ and the spacing between points $\Delta r=2.5\cdot10^{-3}$ (400 points), which gives a Courant factor ($=\case{\Delta t}{\Delta r}$) of $0.2$. For the convergence runs, the lowest resolution had $\Delta t=9\cdot10^{-5}$ and $\Delta r=8.3\cdot10^{-4}$ (1200 points, Courant factor $=0.108$) and the resolution increase for the higher resolutions was a factor of 1.5 in both spatial and time steps.

The parameters for the initial data perturbations (see subsections \ref{initialgauge} and \ref{initialwave}) were $A_{\alpha}=0.1$ in $\alpha$ for the gauge waves and $A_{\Phi}=0.058$ in $\Phi$ for the scalar field evolution, with $\sigma=0.1$ and centered at $c=0.25$ for both. In the scalar field case, an initial perturbation of amplitude $A_{\Phi}=0.06$ in $\Phi$ is enough to collapse to a black hole when it reaches the origin. 

\subsection{Evolution results}

\begin{figure}[htbp!!]
\center
\begin{tabular}{ @{}m{0.06\linewidth} m{0.45\linewidth}@{} @{}m{0.45\linewidth}@{} }
 & \begin{tikzpicture}[scale=1.2]\draw (-1cm,0cm) node {};
		\draw (0cm, 0cm) node {$\chi$}; \draw (0.3cm, 0cm) -- (1cm, 0cm); \draw (1.5cm, 0cm) node {$\gamma_{rr}$}; \draw [dashed] (1.8cm, 0cm) -- (2.5cm, 0cm);
		\draw (3cm, 0cm) node {$\alpha_{unity}$}; \draw [dotted] (3.3cm, 0cm) -- (4cm, 0cm);
	\end{tikzpicture}
&	\begin{tikzpicture}[scale=1.2]\draw (-1cm,0cm) node {};
		\draw (0cm, 0cm) node {$A_{rr}$}; \draw (0.3cm, 0cm) -- (1cm, 0cm); \draw (1.5cm, 0cm) node {$\K$}; \draw [dashed] (1.8cm, 0cm) -- (2.5cm, 0cm);
		\draw (3cm, 0cm) node {$\Lambda^r$}; \draw [dotted] (3.3cm, 0cm) -- (4cm, 0cm); 
	\end{tikzpicture}
\\
\center Time = 0.00 &
\mbox{\includegraphics[width=1\linewidth]{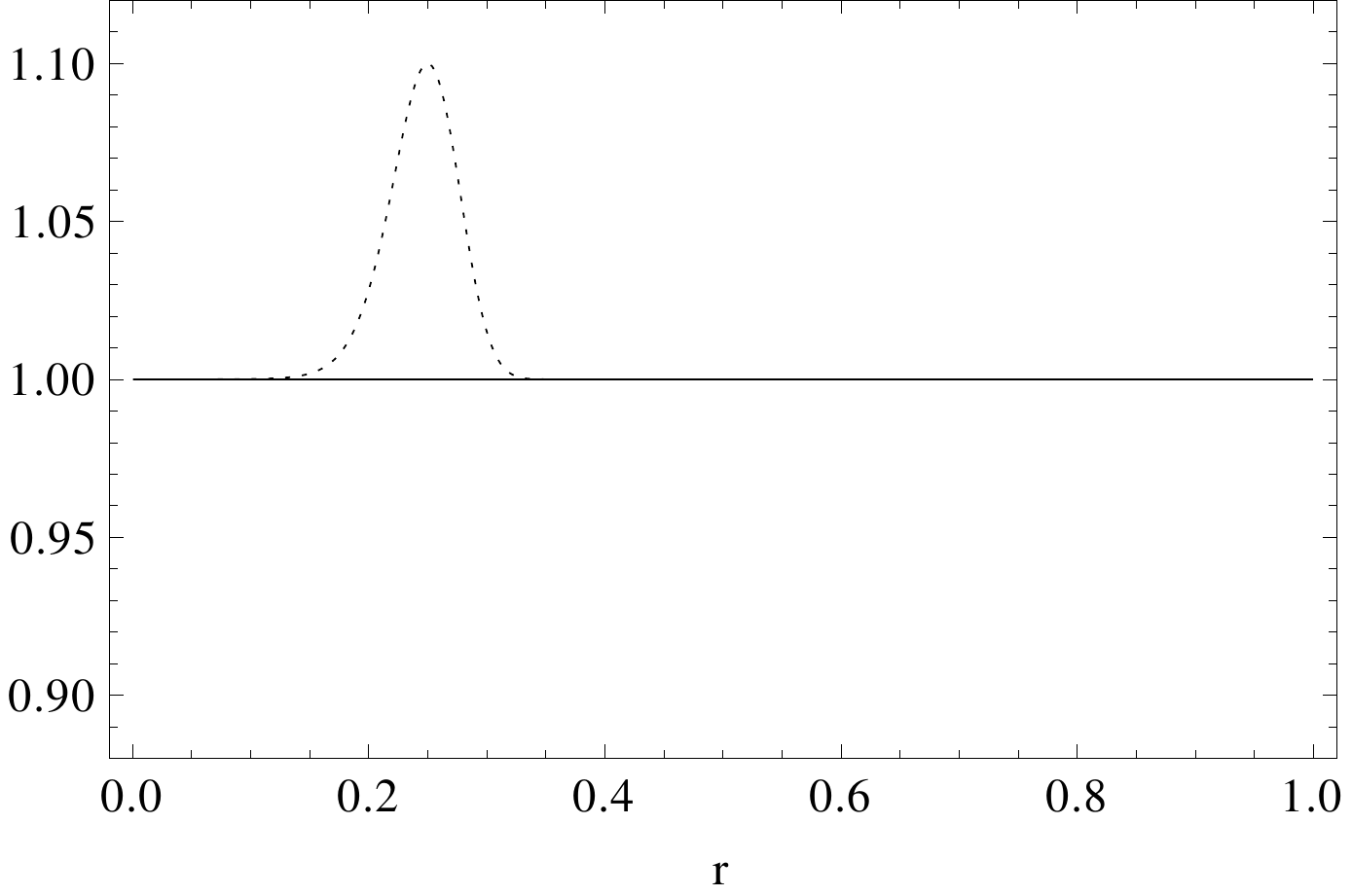}}&
\hspace{-1.0ex} \mbox{\includegraphics[width=1\linewidth]{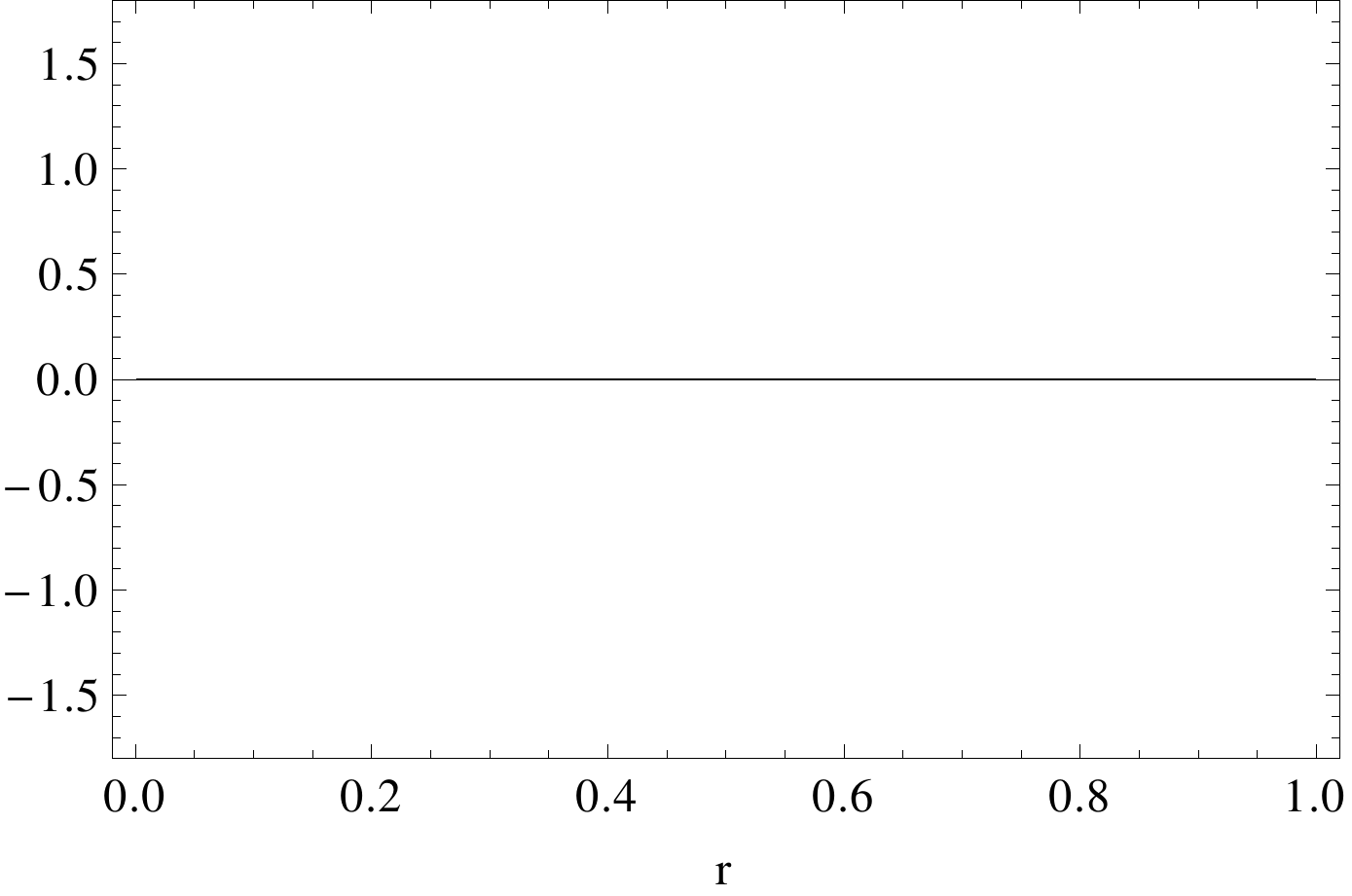}}\\
\center Time = 0.40 &
\vspace{-4.9ex} \mbox{\includegraphics[width=1\linewidth]{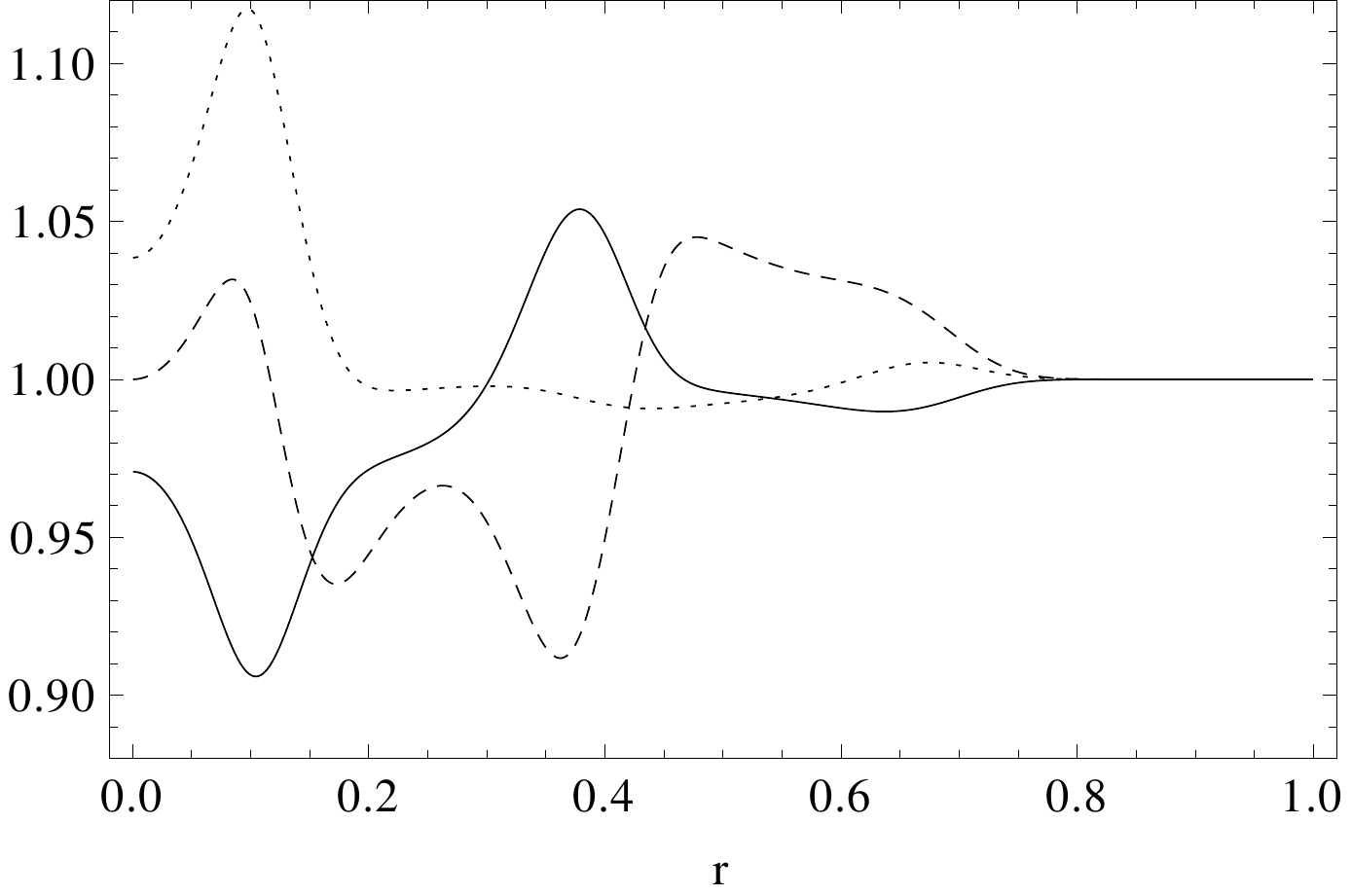}}&
\vspace{-4.9ex} \hspace{-1.0ex} \mbox{\includegraphics[width=1\linewidth]{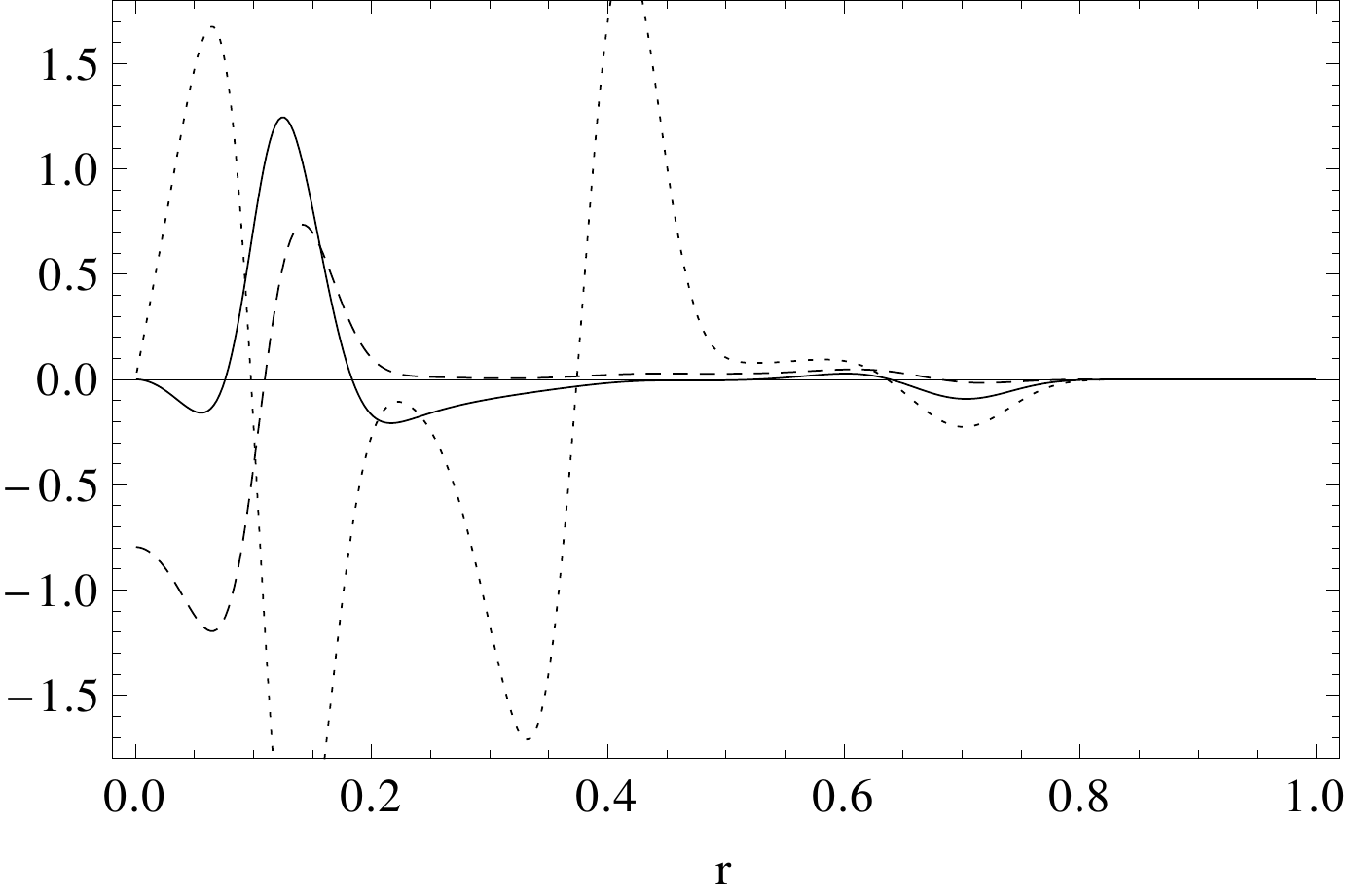}}\\
\center Time = 0.90 &
\vspace{-4.9ex} \mbox{\includegraphics[width=1\linewidth]{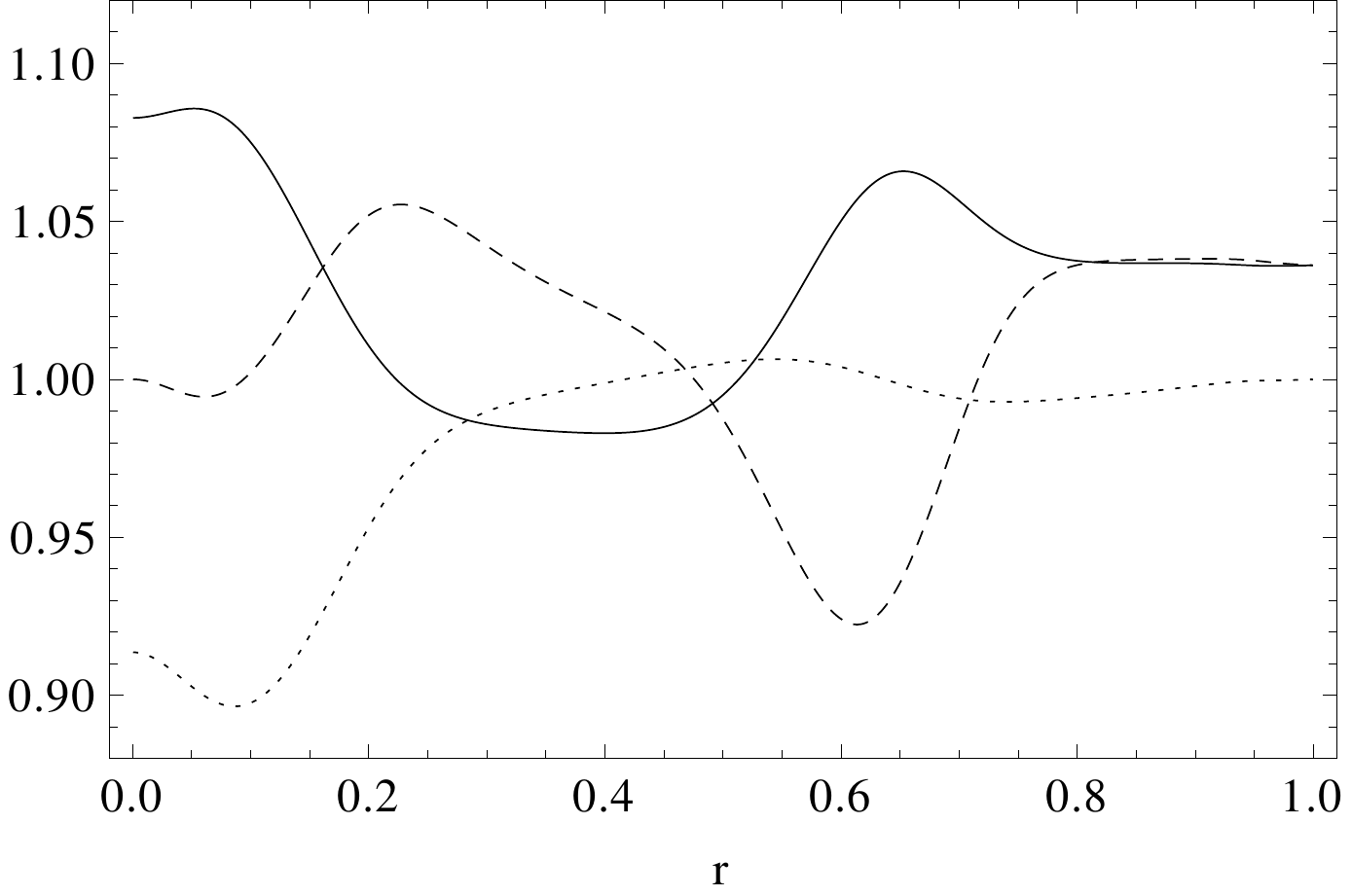}}&
\vspace{-4.9ex} \hspace{-1.0ex} \mbox{\includegraphics[width=1\linewidth]{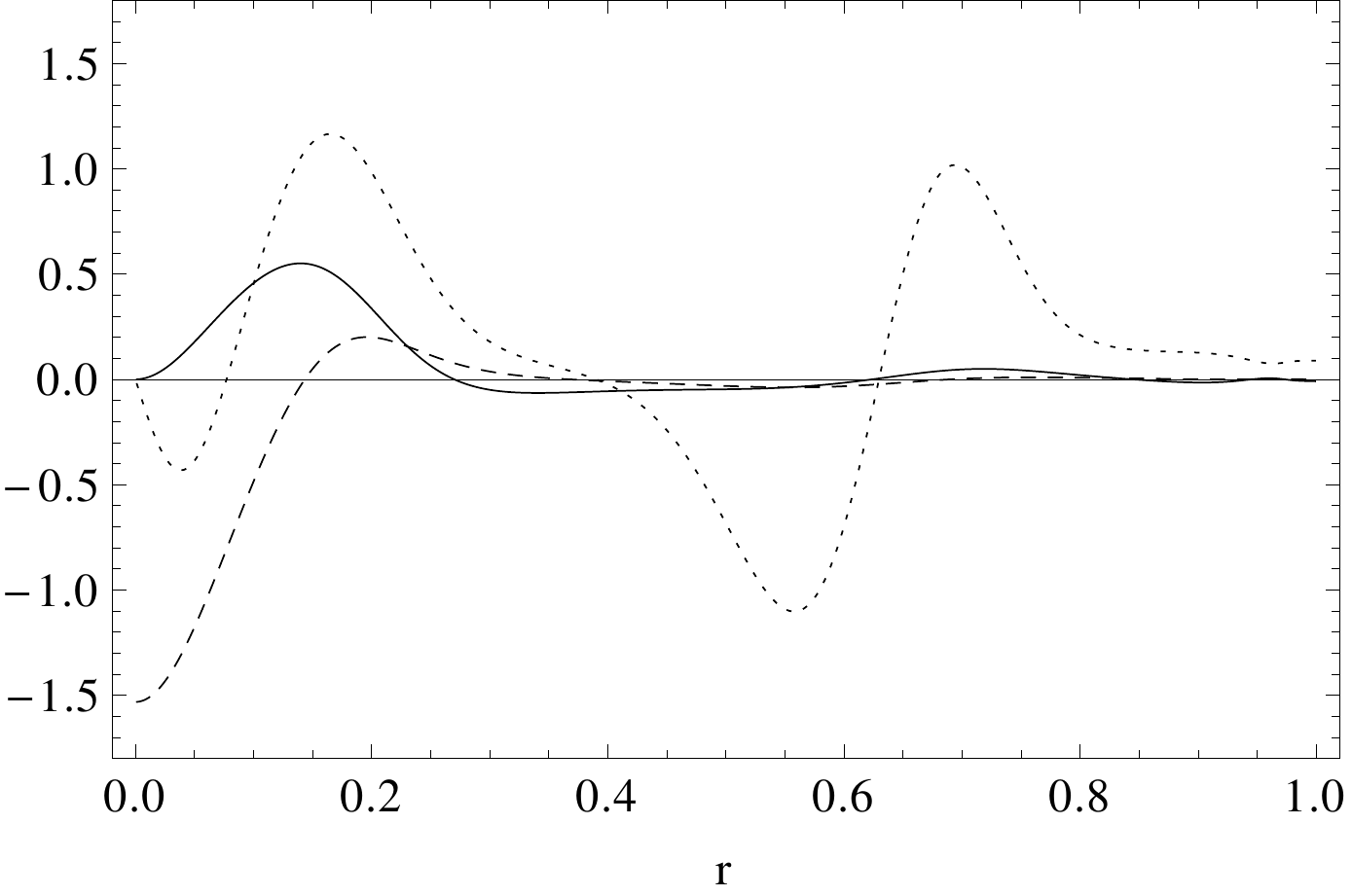}}\\
\center Time = 1.40 &
\vspace{-4.9ex} \mbox{\includegraphics[width=1\linewidth]{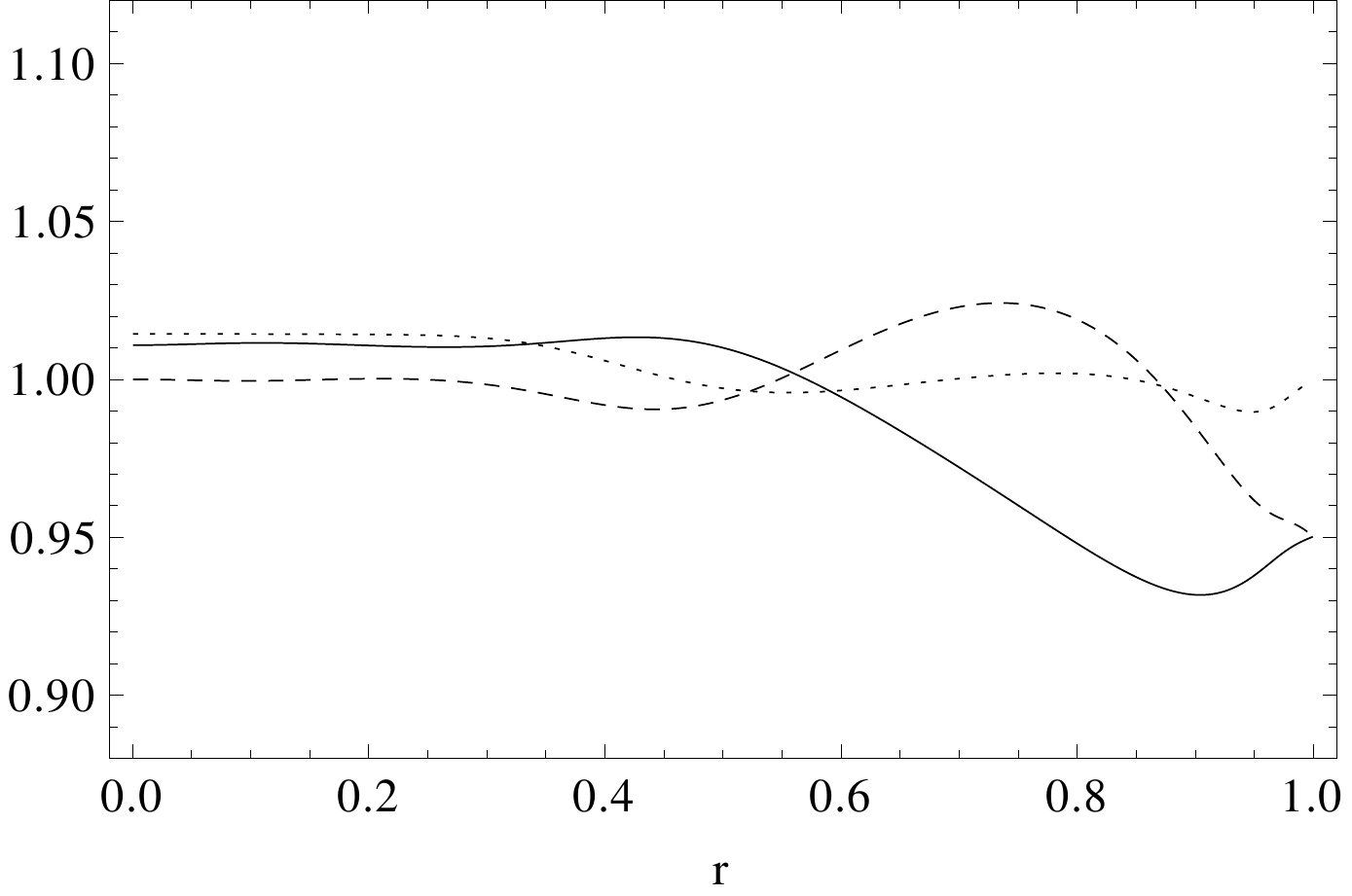}}&
\vspace{-4.9ex} \hspace{-1.0ex} \mbox{\includegraphics[width=1\linewidth]{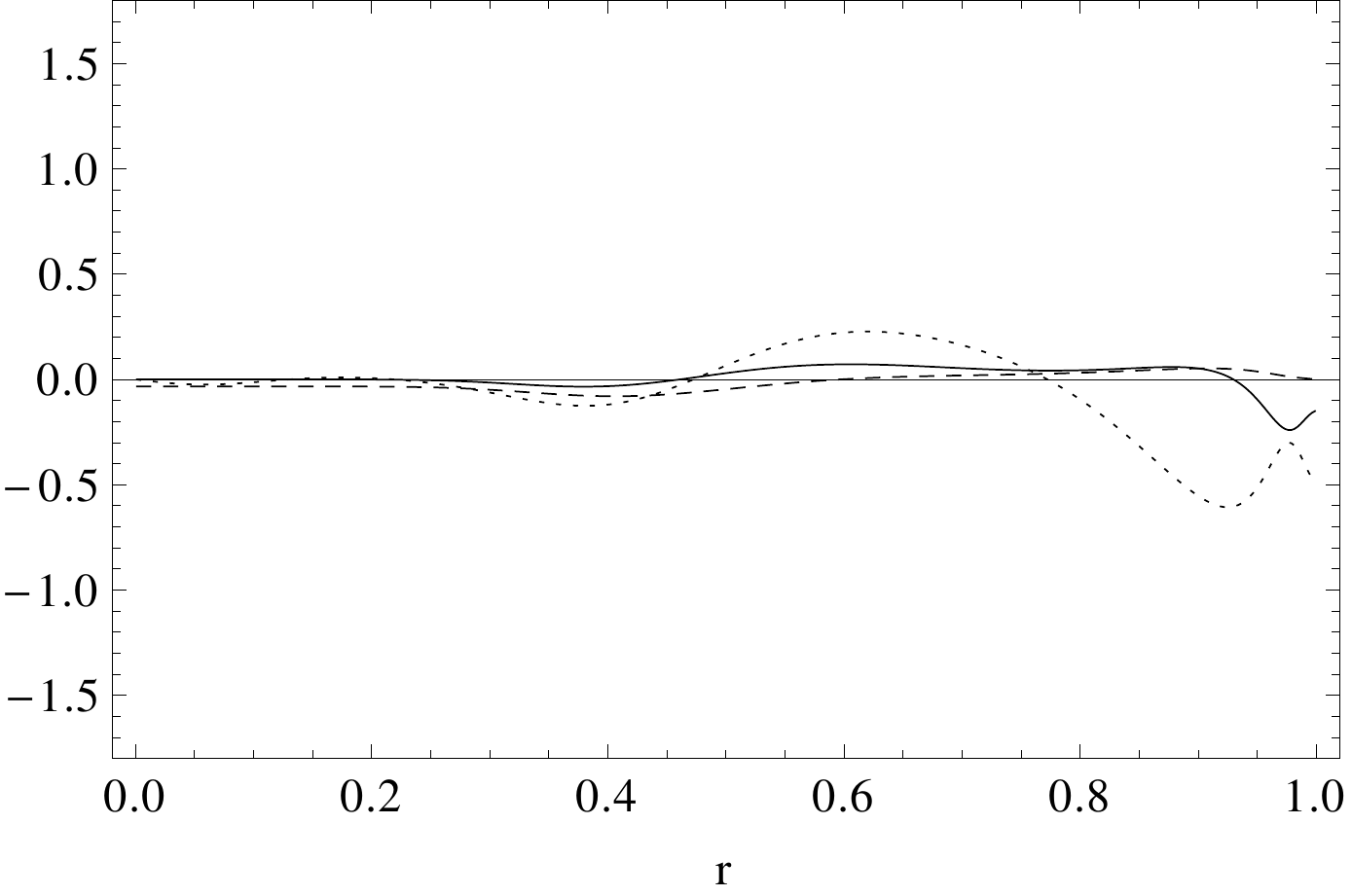}}\\
\center Time = 3.00 &
\vspace{-4.9ex} \mbox{\includegraphics[width=1\linewidth]{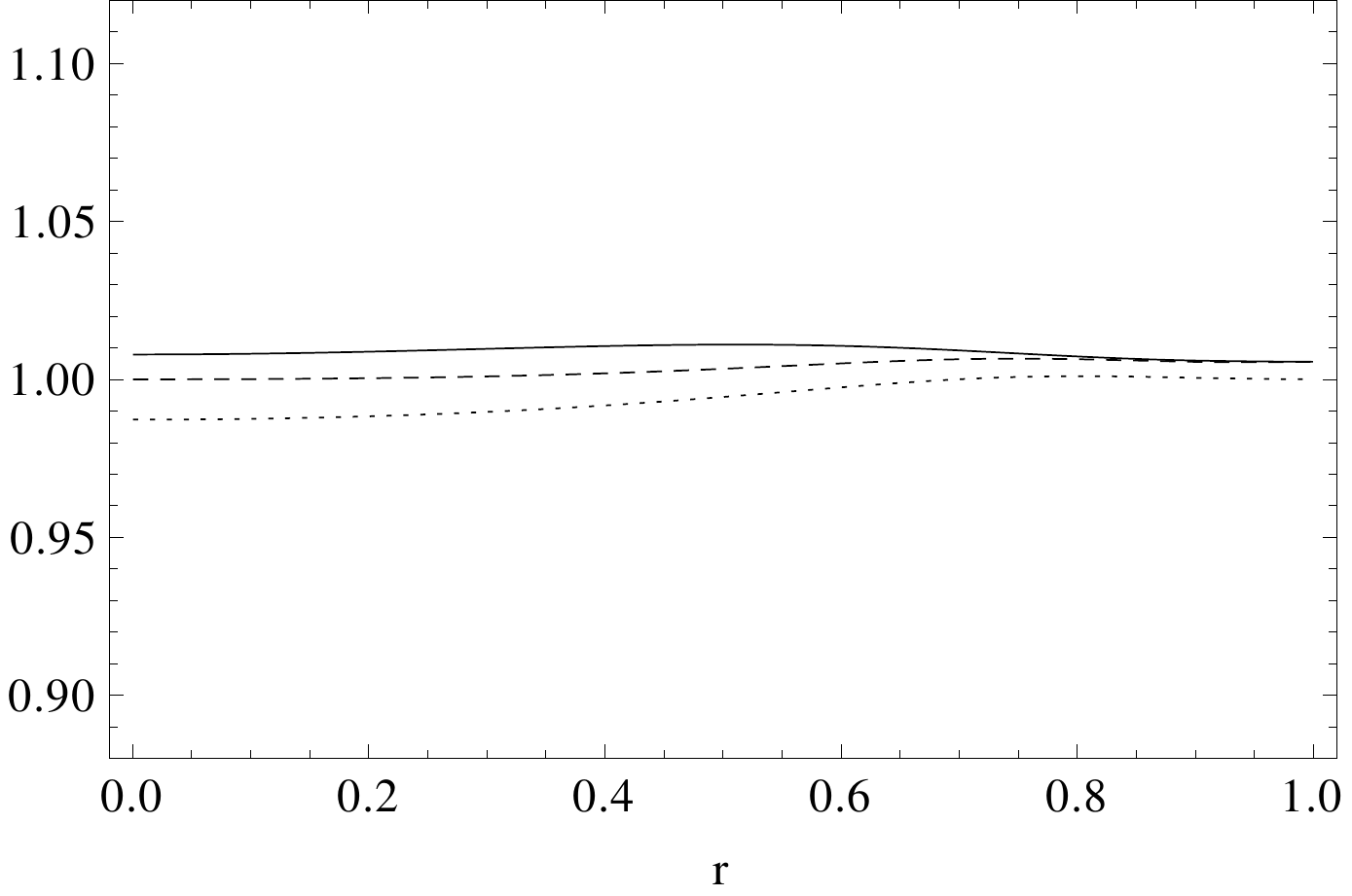}}&
\vspace{-4.9ex} \hspace{-1.0ex} \mbox{\includegraphics[width=1\linewidth]{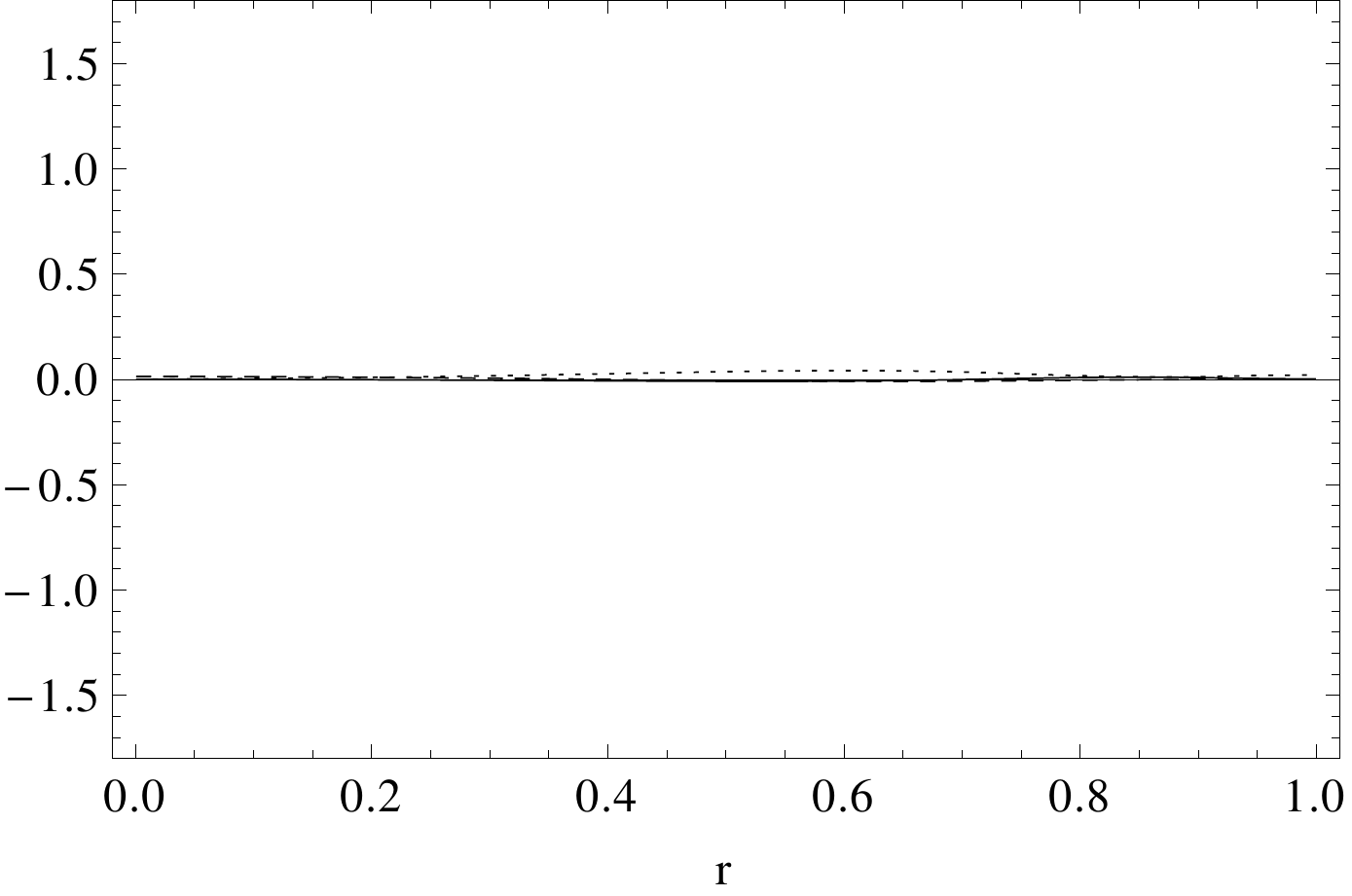}}
\end{tabular}
\caption{Evolution of the variables (gauge waves).}
\label{allevolgauge}
\end{figure}
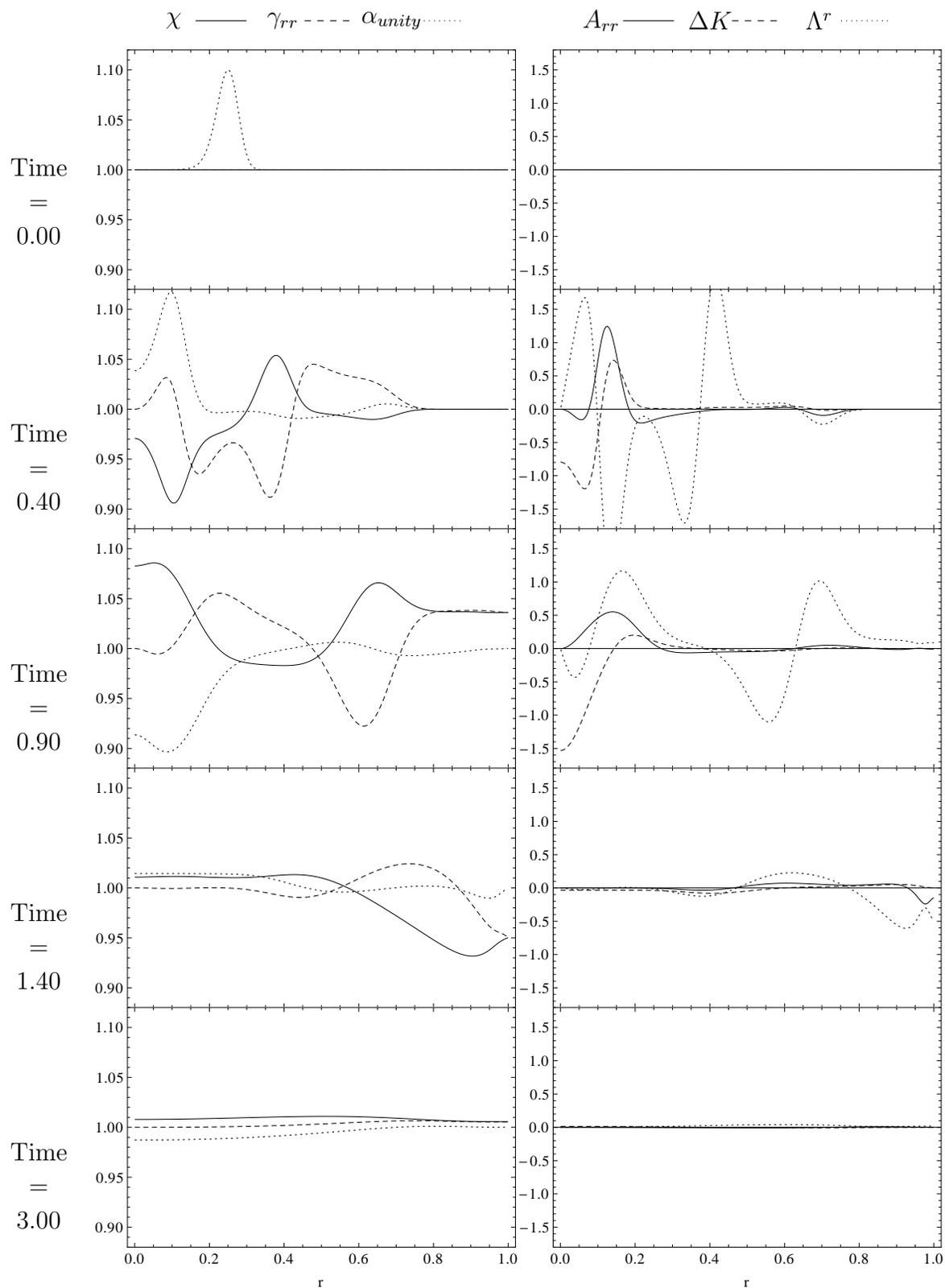

\begin{figure}[htbp!!]
\center
\begin{tabular}{ @{}m{0.06\linewidth} m{0.45\linewidth}@{} @{}m{0.45\linewidth}@{} }
 & \begin{tikzpicture}[scale=1.2] \draw (-1.3cm,0cm) node {};
		\draw (0cm, 0cm) node {$\chi$}; \draw (0.3cm, 0cm) -- (1cm, 0cm); \draw (1.5cm, 0cm) node {$\gamma_{rr}$}; \draw [dashed] (1.8cm, 0cm) -- (2.5cm, 0cm);
		\draw (3cm, 0cm) node {$\alpha_{unity}$}; \draw [dotted] (3.3cm, 0cm) -- (4cm, 0cm); 
	\end{tikzpicture}
&	\begin{tikzpicture}[scale=0.9] \draw (-0.5cm,0cm) node {};
		\draw (0cm, 0cm) node {$A_{rr}$}; \draw (0.3cm, 0cm) -- (1cm, 0cm); \draw (1.5cm, 0cm) node {$\K$}; \draw [dashed] (1.8cm, 0cm) -- (2.5cm, 0cm);
		\draw (3cm, 0cm) node {$\Lambda^r$}; \draw [dotted] (3.3cm, 0cm) -- (4cm, 0cm); \draw (4.5cm, 0cm) node {$\case{\Phi}{\Omega}$}; \draw [dash pattern= on 4pt off 2pt on 1pt off 2pt] (4.8cm, 0cm) -- (5.5cm, 0cm);
		\draw (6cm, 0cm) node {$\case{\Pi}{\Omega}$}; \draw [dash pattern= on 8pt off 2pt] (6.3cm, 0cm) -- (7cm, 0cm);
	\end{tikzpicture}
\\
\center Time = 0.00 &
\mbox{\includegraphics[width=1\linewidth]{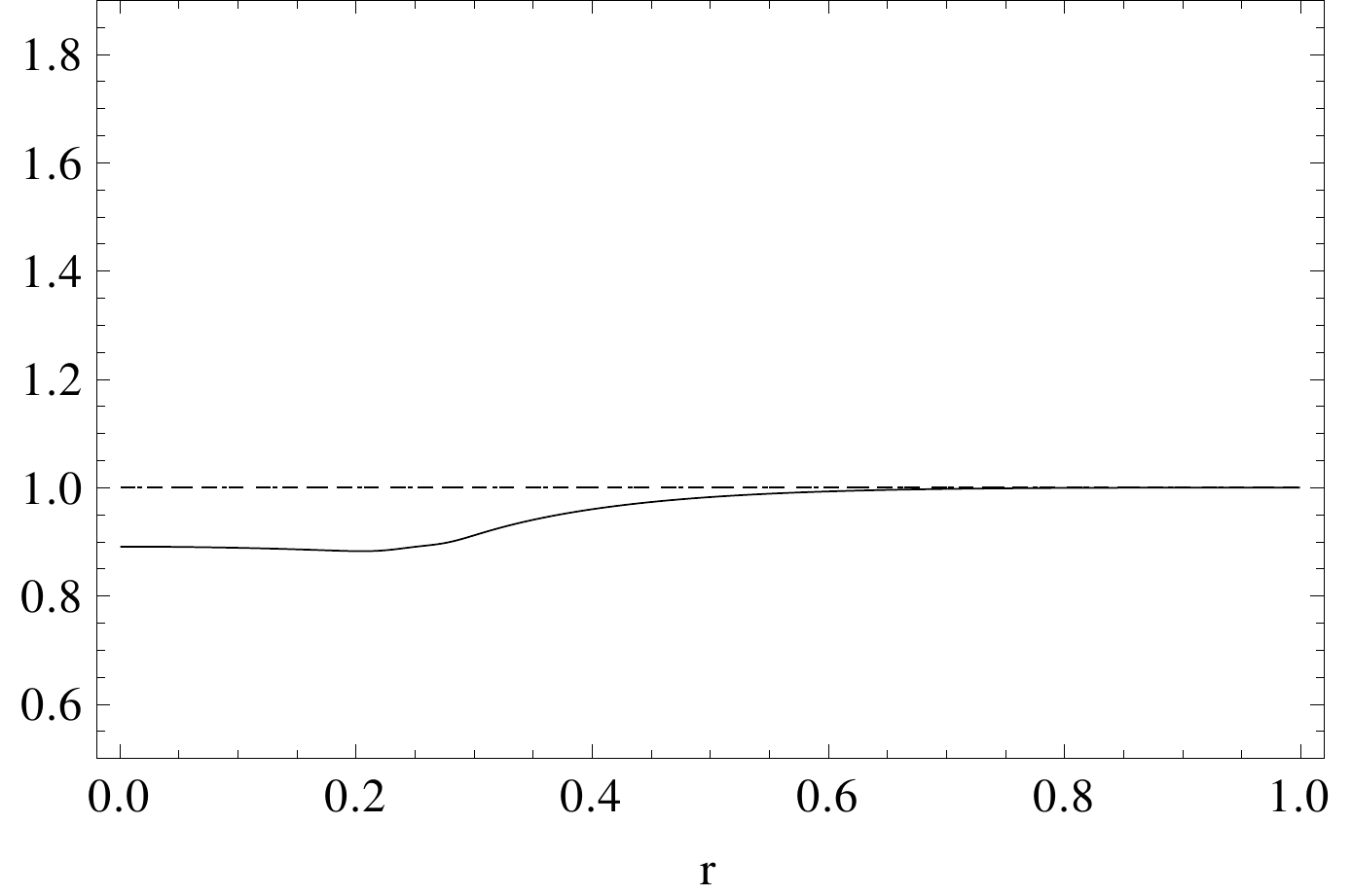}}&
\hspace{-1.0ex} \mbox{\includegraphics[width=1\linewidth]{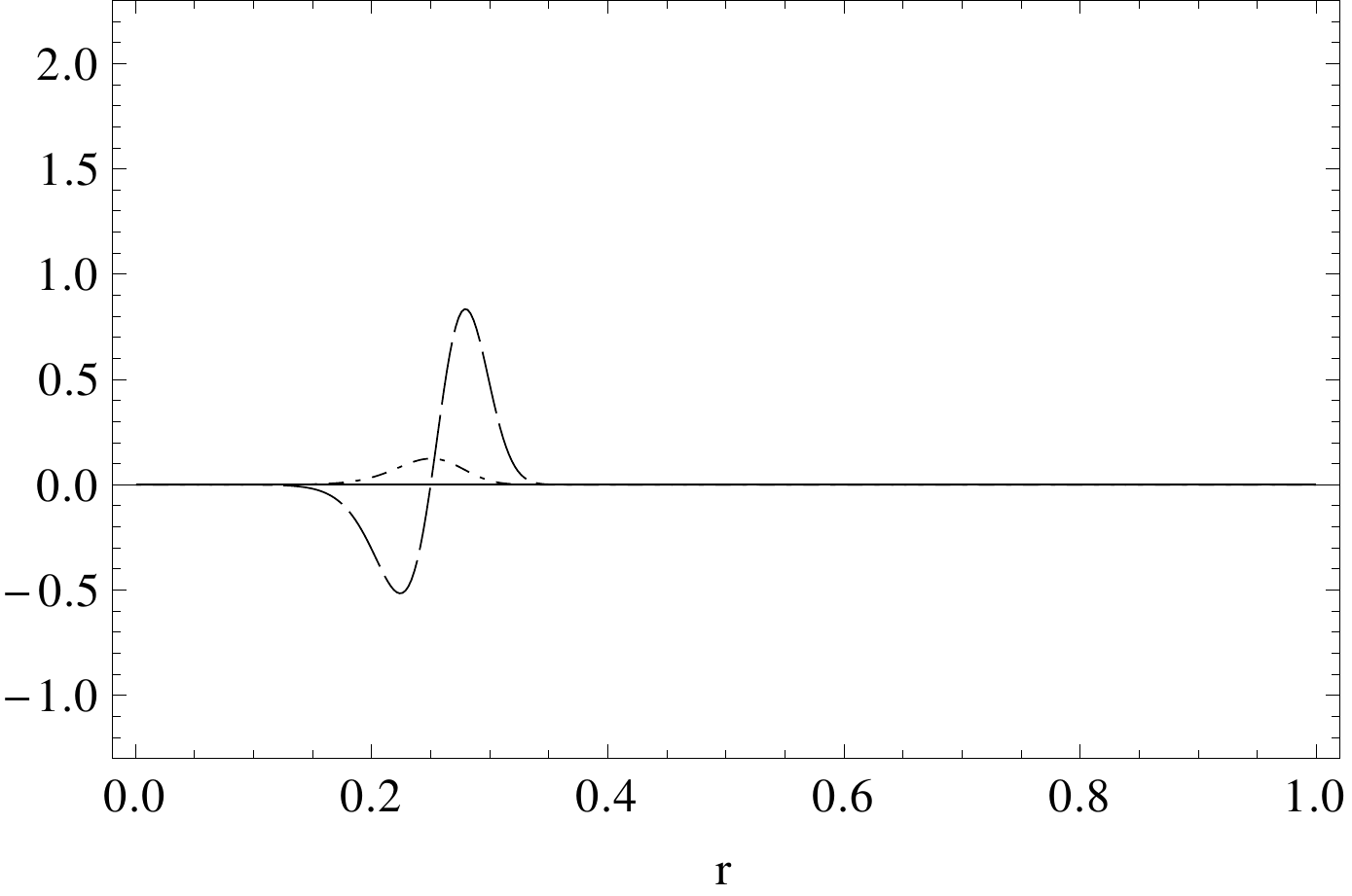}}\\
\center Time = 0.40 &
\vspace{-4.9ex} \mbox{\includegraphics[width=1\linewidth]{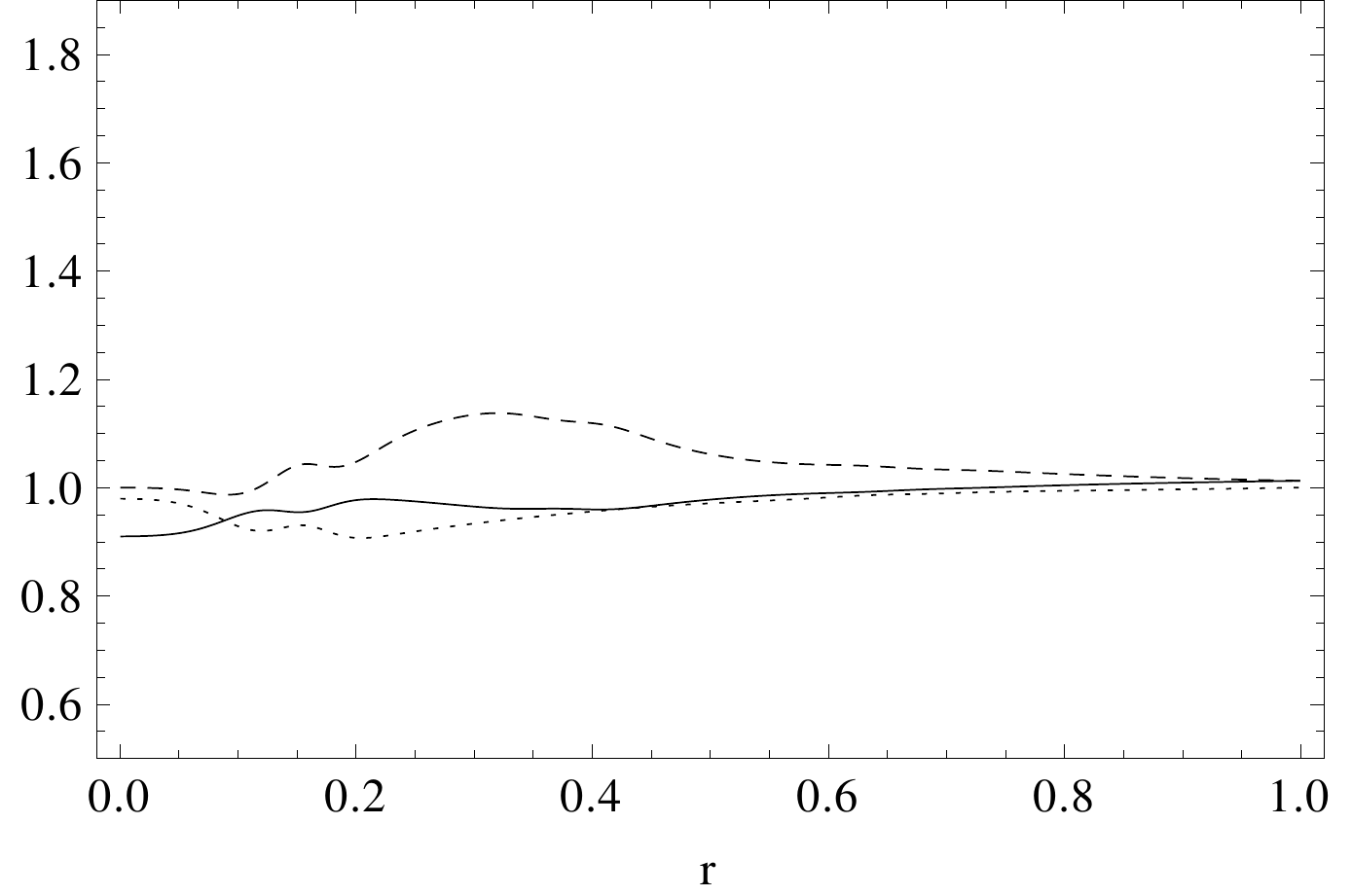}}&
\vspace{-4.9ex} \hspace{-1.0ex} \mbox{\includegraphics[width=1\linewidth]{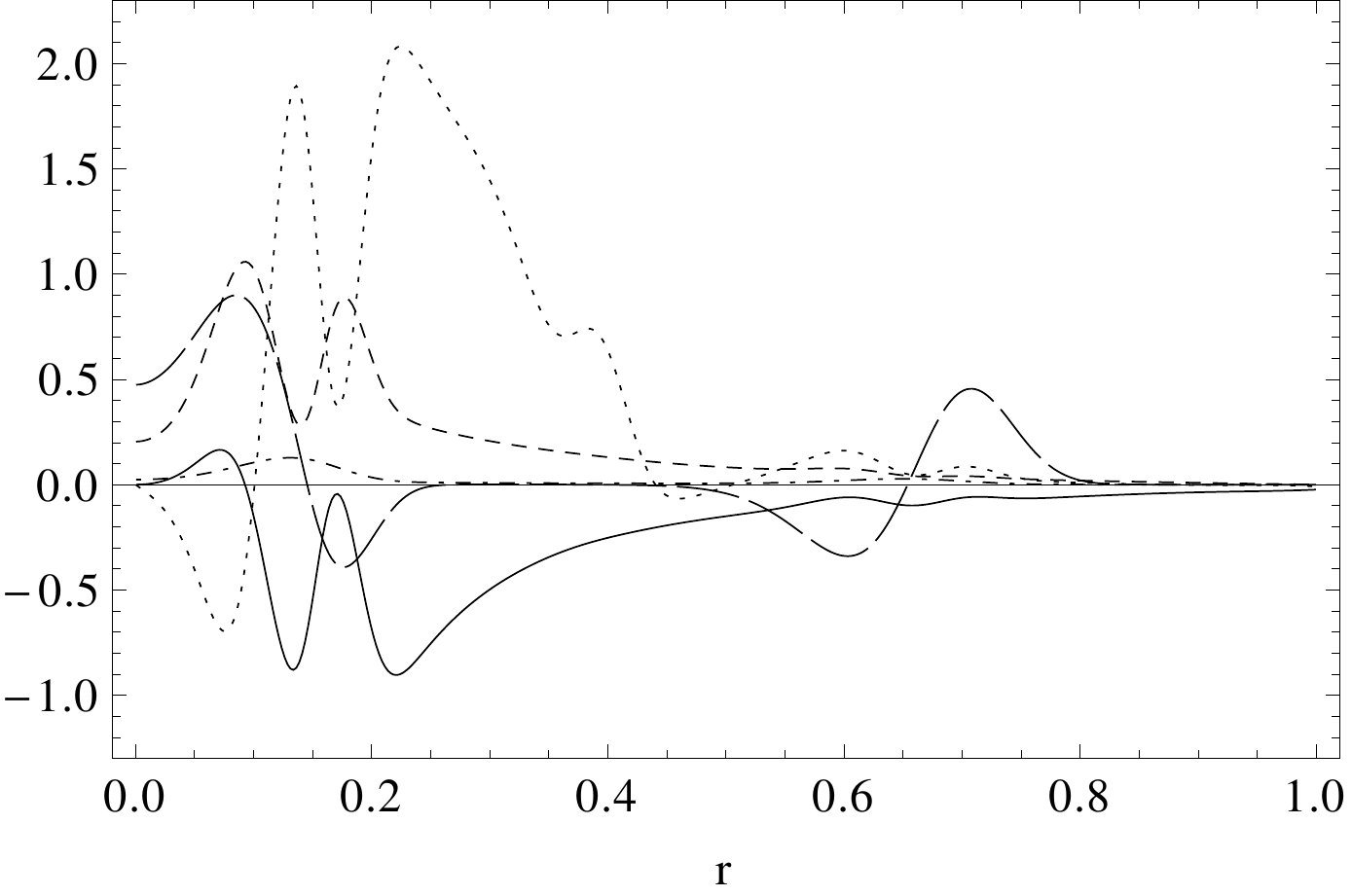}}\\
\center Time = 1.50 &
\vspace{-4.9ex} \mbox{\includegraphics[width=1\linewidth]{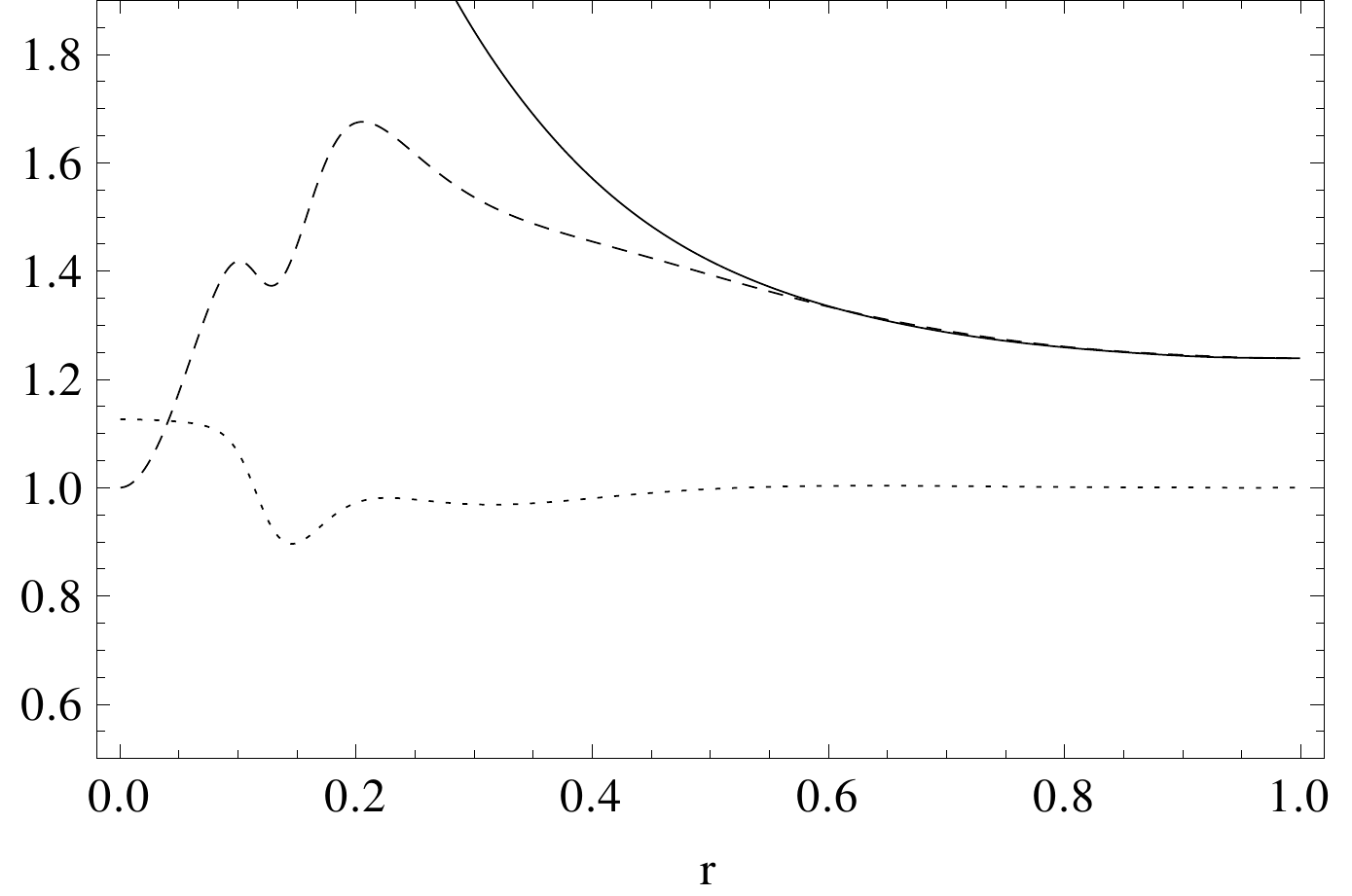}}&
\vspace{-4.9ex} \hspace{-1.0ex} \mbox{\includegraphics[width=1\linewidth]{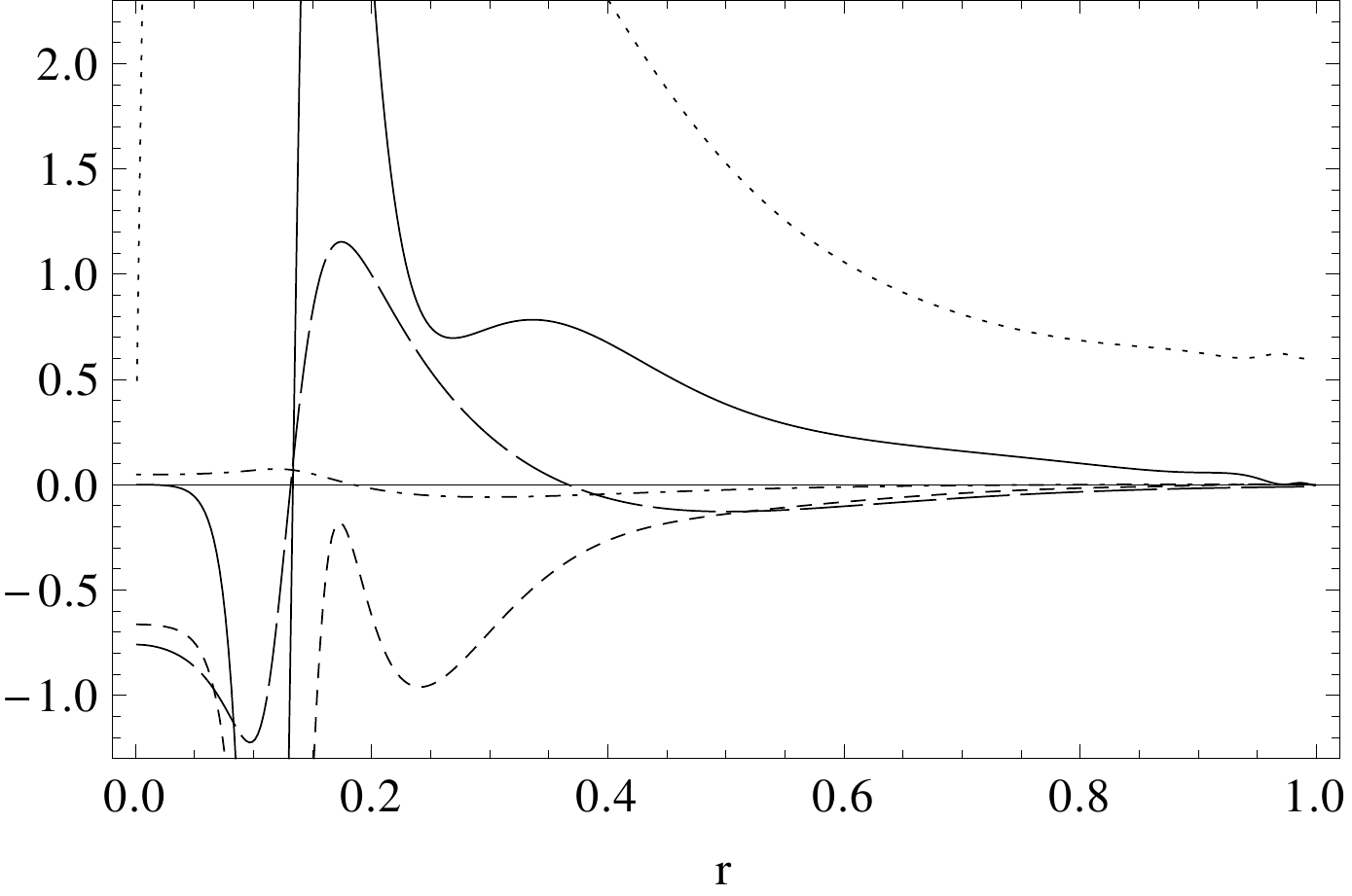}}\\
\center Time = 3.60 &
\vspace{-4.9ex} \mbox{\includegraphics[width=1\linewidth]{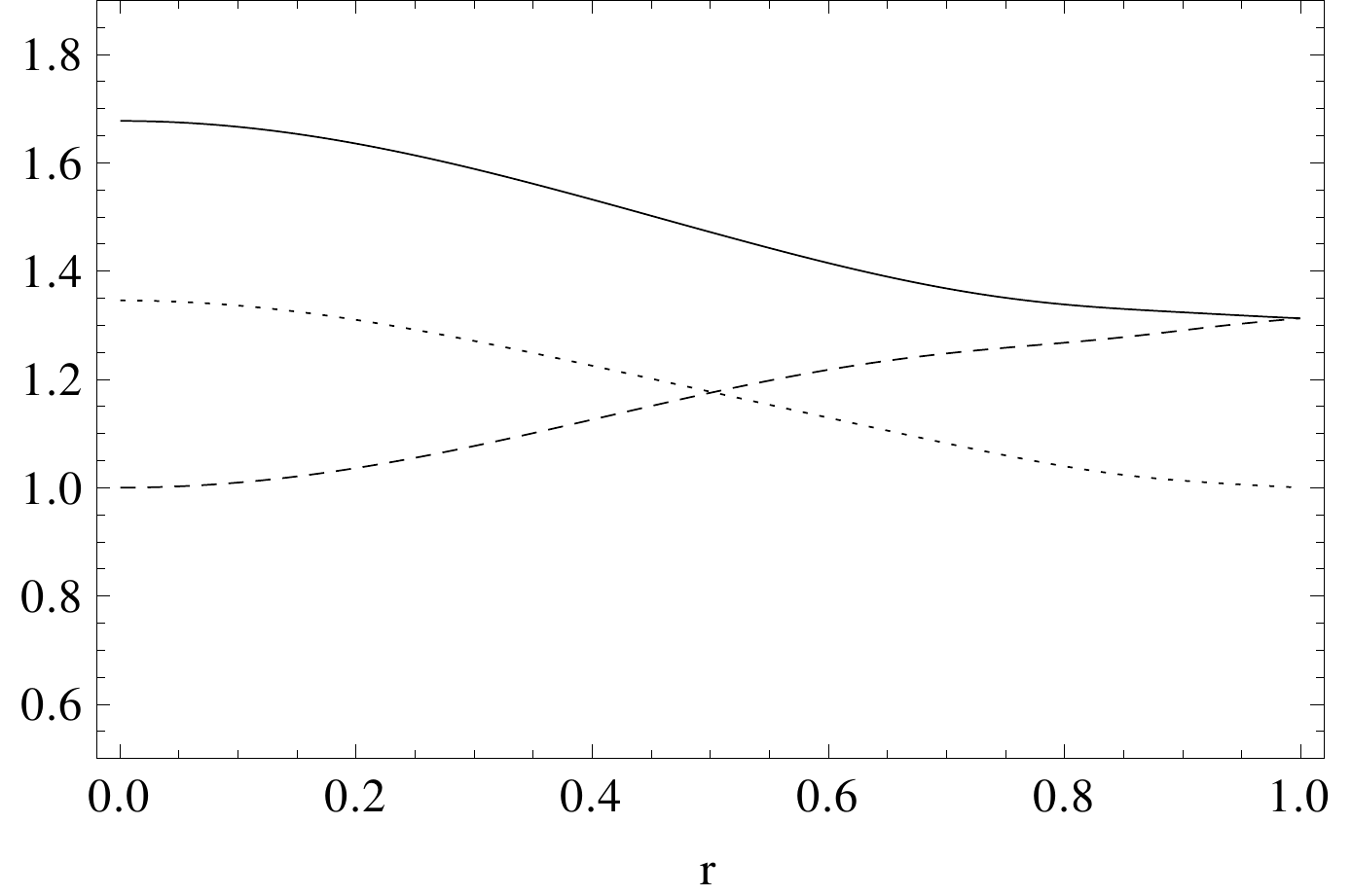}}&
\vspace{-4.9ex} \hspace{-1.0ex} \mbox{\includegraphics[width=1\linewidth]{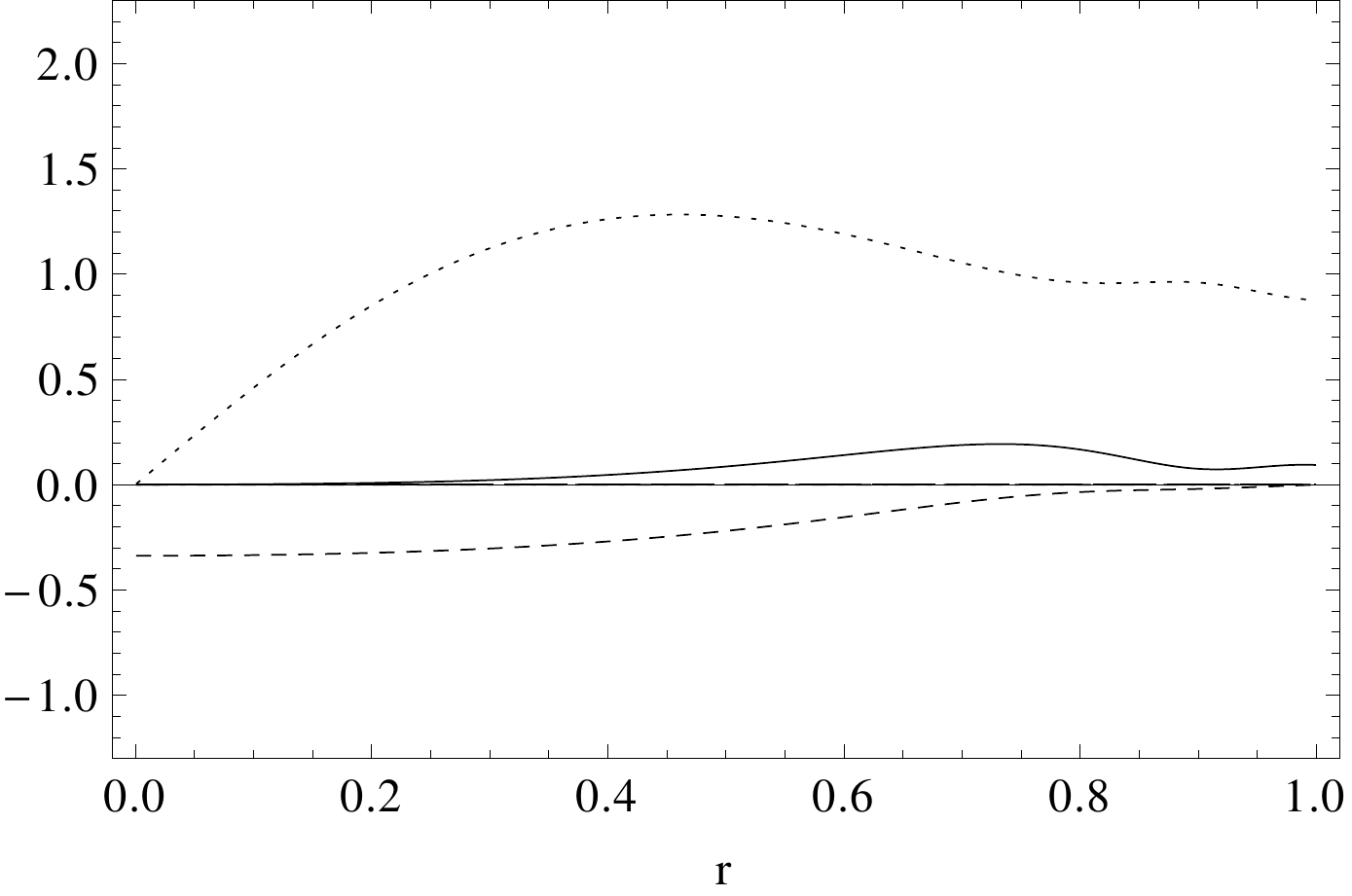}}\\
\center Time = 6.00 &
\vspace{-4.9ex} \mbox{\includegraphics[width=1\linewidth]{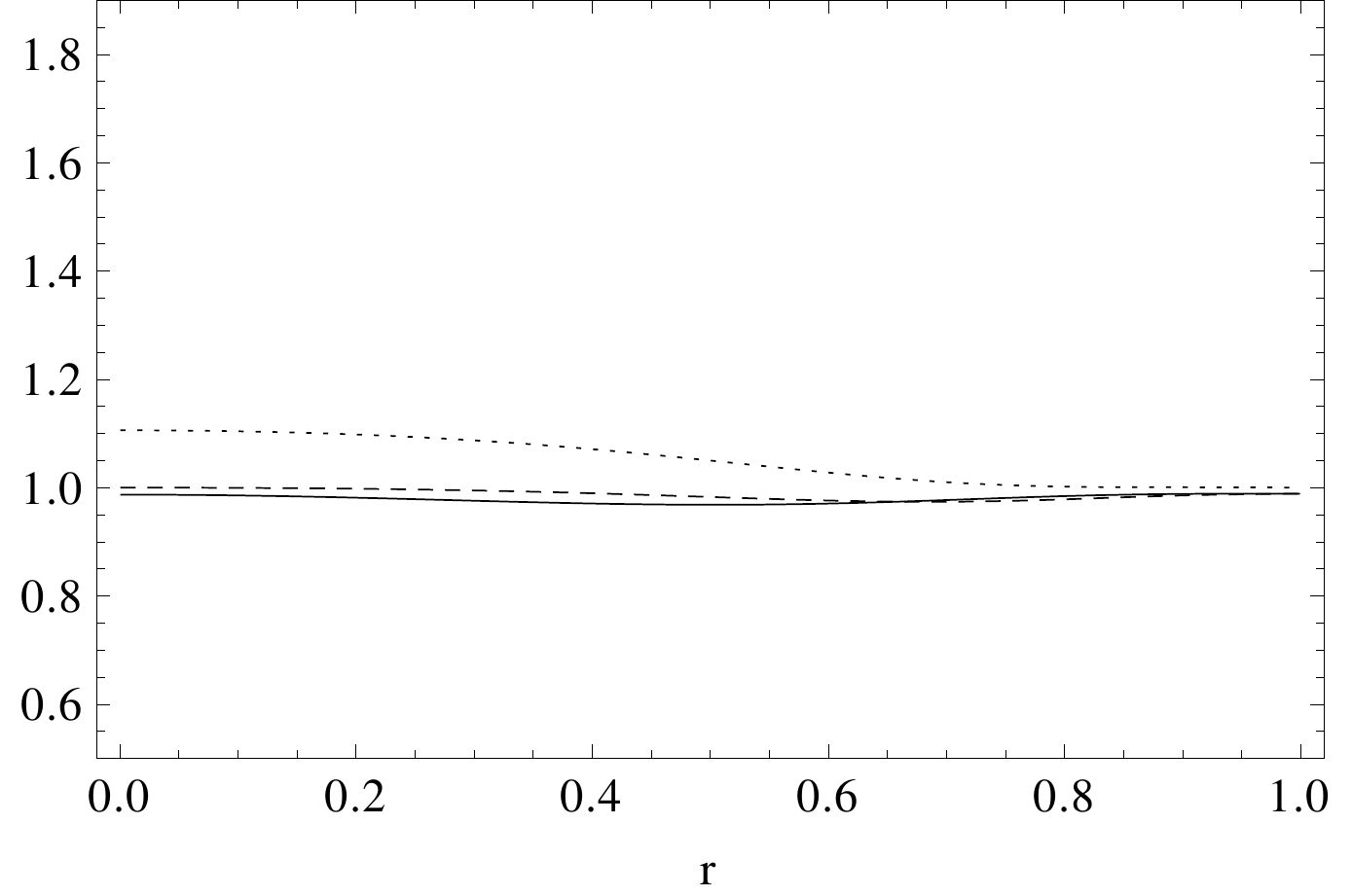}}&
\vspace{-4.9ex} \hspace{-1.0ex} \mbox{\includegraphics[width=1\linewidth]{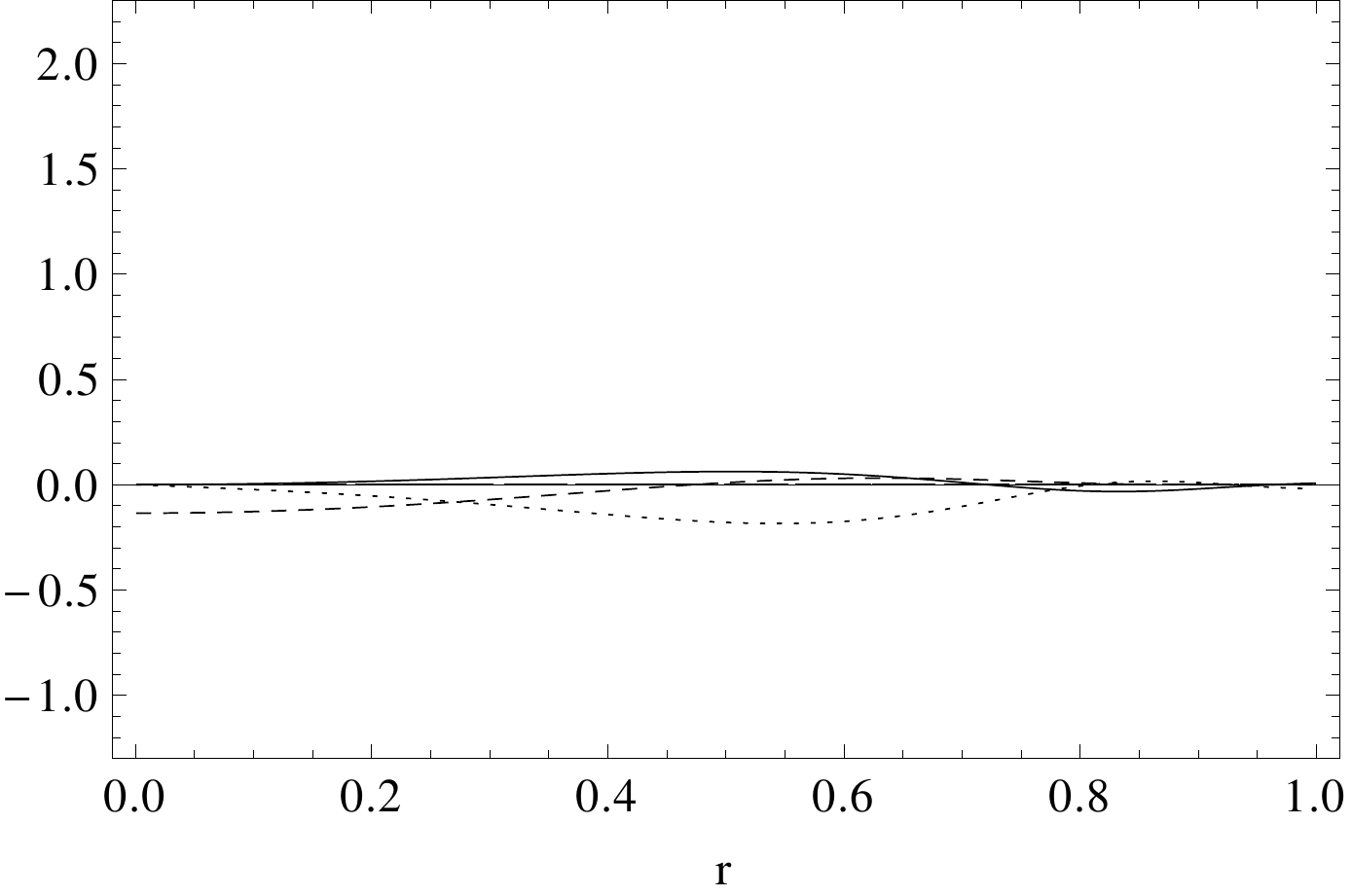}}
\end{tabular}
\caption{Evolution of the variables (scalar field).}
\label{allevolkgf}
\end{figure}

In figures \ref{allevolgauge} and \ref{allevolkgf} we show two simulations performed using the GBSSN equations. The variables as a function of $r$ are displayed at the indicated times. On the left we present the variables that equal 1 in flat spacetime ($\chi$, $\gamma_{rr}$ and $\alpha_{unity}$ - note that $\gamma_{\theta\theta}$ is eliminated in terms of $\gamma_{rr}$) and on the right the ones that are zero in flat spacetime ($A_{rr}$, $\K$, $\Lambda^r$, $\case{\Phi}{\Omega}$ and $\case{\Pi}{\Omega}$). 
The figures only show the results up to $t=6$, but the stability of the simulations has been checked until $t=10\,000$ for the scalar field evolution; from $t\approx250$ on, only noise is seen in the evolution variables, so that we conclude that the simulation can run forever. 
The gauge waves initial data simulation can be seen in \fref{allevolgauge}. The initial perturbation in the lapse excites the rest of the variables and splits into two pulses: the out-going one moves towards $\scri^+$ (located at $r=1$) and leaves, while the other propagates to the origin, is reflected there and then follows the other pulse and also leaves the integration domain. After a certain time, all perturbations disappear and flat spacetime remains. 

For the simulations with scalar field we first solve the Hamiltonian constraint \eref{inihamil} to obtain the initial data for the conformal factor $\chi$. Our choice corresponds to choosing an isotropic radial coordinate, and we show the initial data in figure \ref{allevolkgf}. Similarly as before, the initial perturbation is reflected at the origin and crosses $\scri^+$ to leave. The other variables react to the perturbation and settle down to their flat spacetime values some time after the scalar field perturbation has left. Note that $\alpha_{unity}$ is always 1 at null infinity, as well as $\chi=\gamma_{rr}$ at that point. 

\begin{figure}[htbp!!]
\center
\mbox{\includegraphics[width=0.65\linewidth]{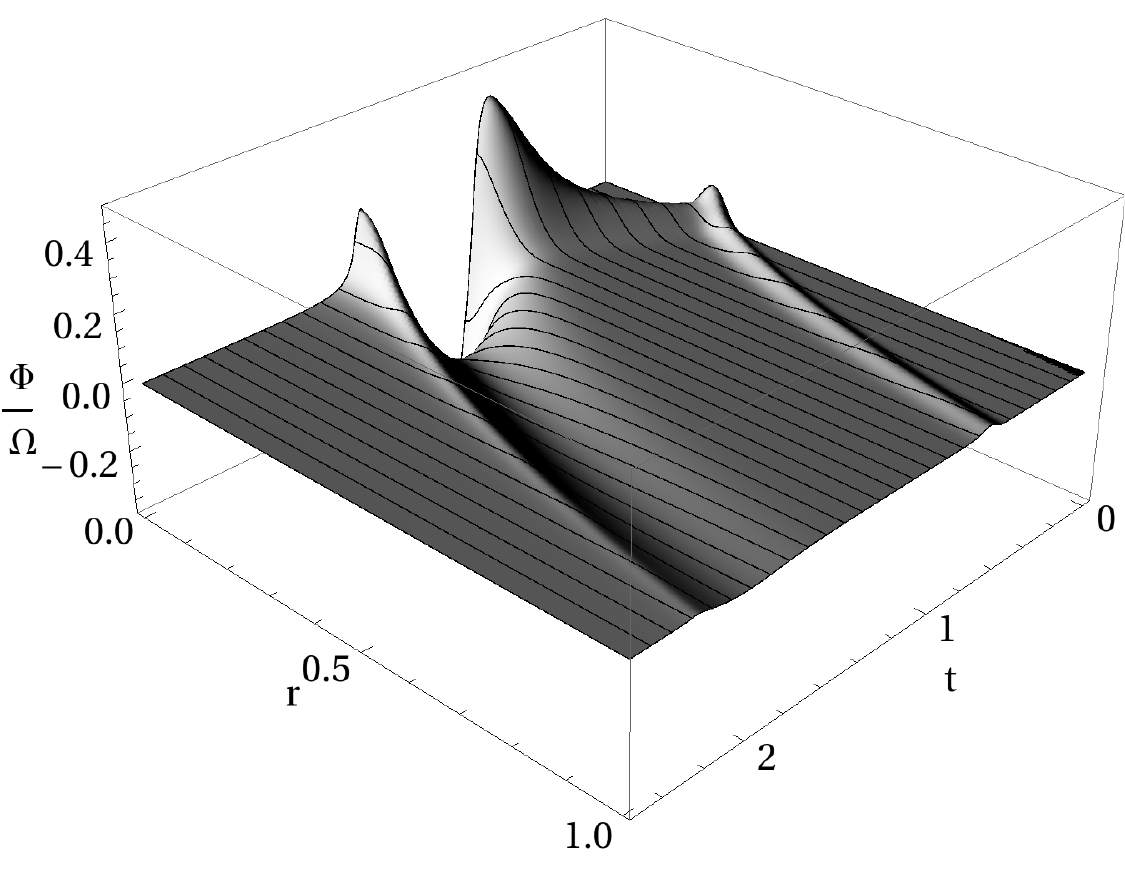}}\vspace{-2ex}
\caption{Behaviour of $\case{\Phi}{\Omega}$ over time and compactified spatial coordinate.}
\label{phirt}
\end{figure}

Figure \ref{phirt} is included to show in more detail the behaviour of the rescaled scalar field $\case{\Phi}{\Omega}$. Here we can see clearly the splitting of the initial perturbation and the fact that the propagation speed neither diverges nor becomes zero, but remains constant. 

\begin{figure}[htbp!!]
\center
\begin{tabular}{@{}c@{}@{}c@{}}
\mbox{\includegraphics[width=0.54\linewidth]{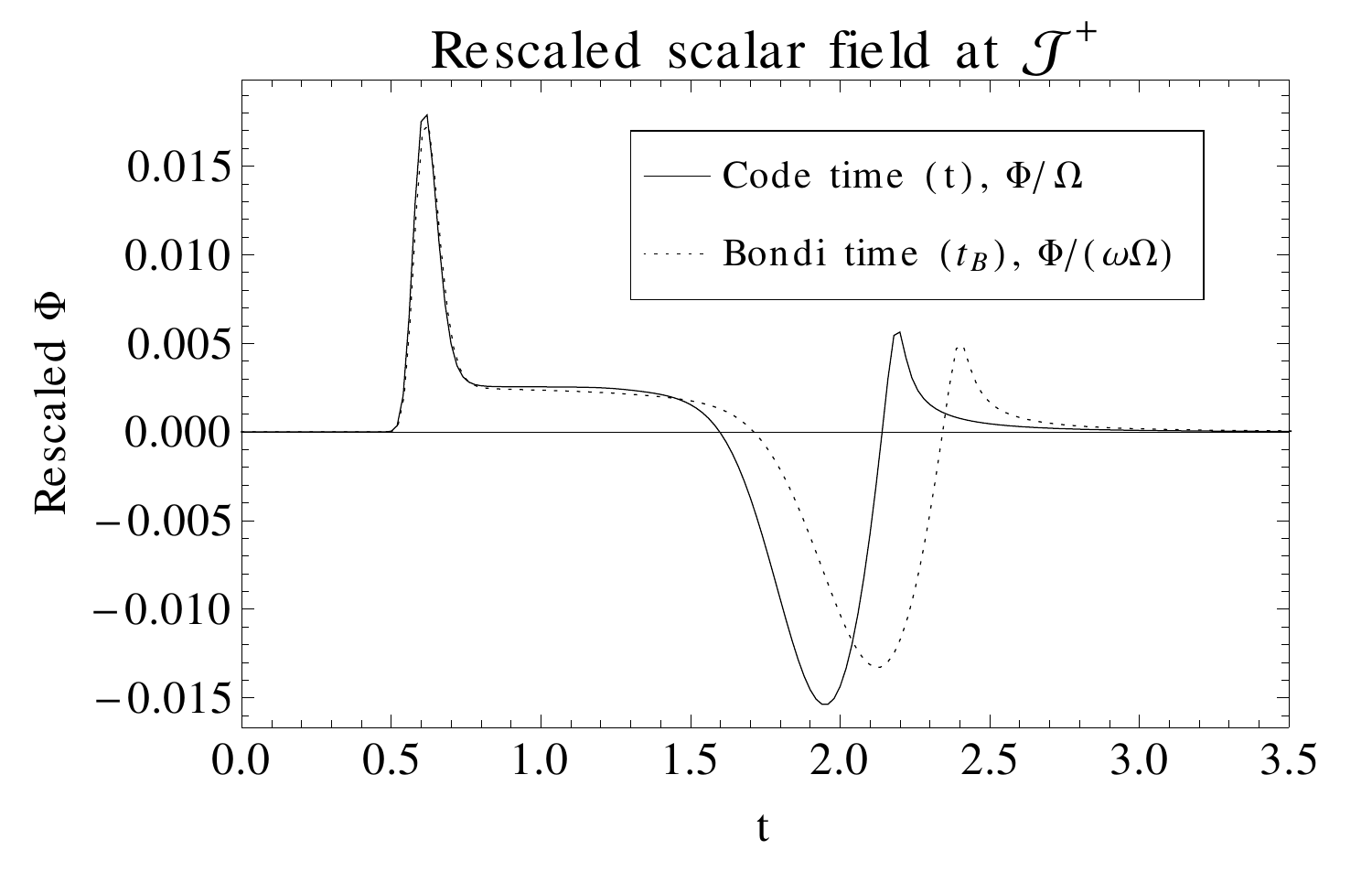}}& 
\hspace{-2ex}\mbox{\includegraphics[width=0.54\linewidth]{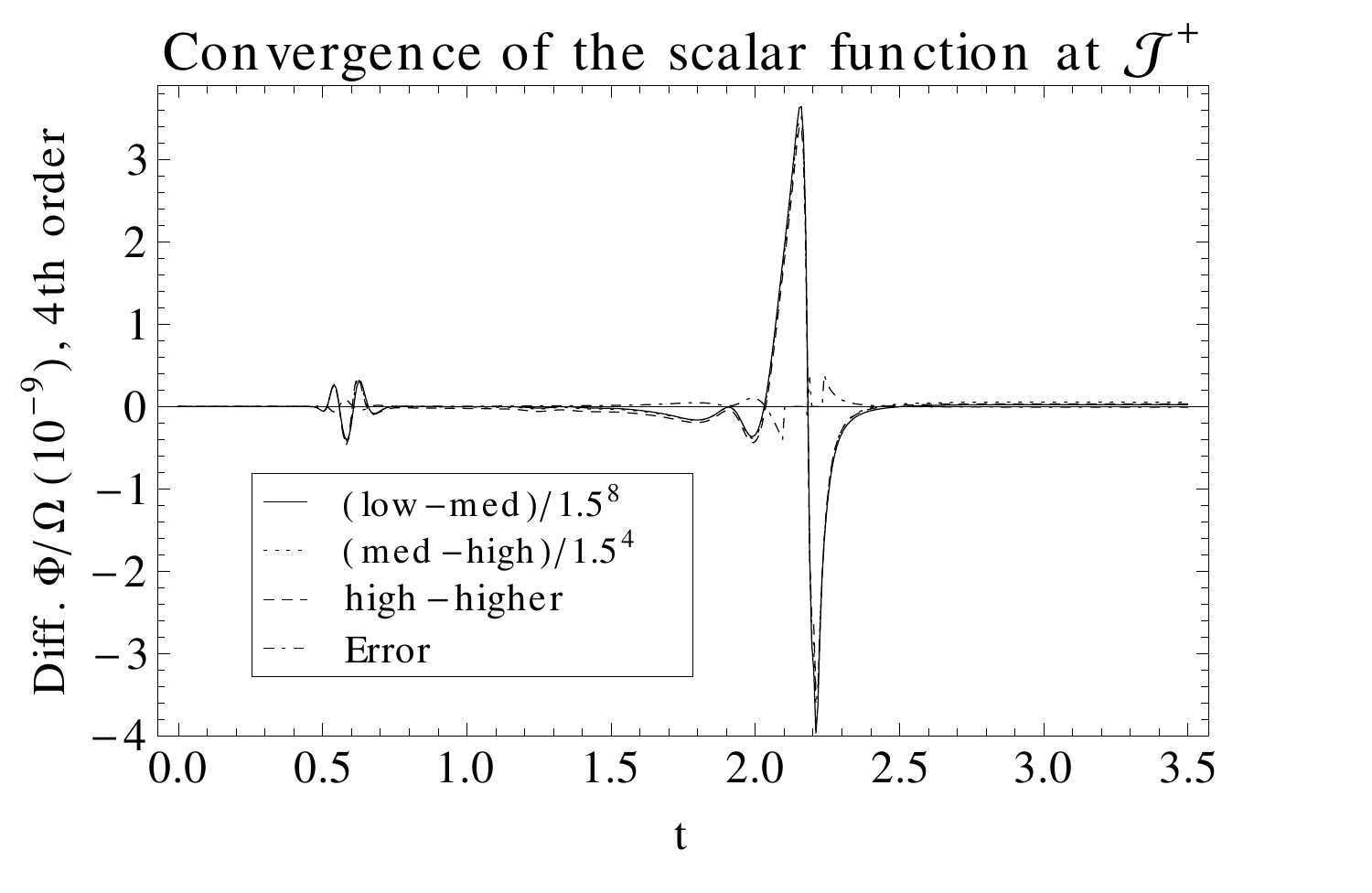}}
\end{tabular}\vspace{-3ex}
\caption{Signal of $\case{\Phi}{\Omega}$ over time at $\scri^+$ on the left and its convergence on the right.}
\label{phiscri}
\end{figure}

On the left in figure \ref{phiscri} we can see the profile of the scalar field signal extracted at future null infinity. The first outgoing pulse and the one reflected at the origin can be clearly seen. The simulation was carried out using a staggered grid, so that extrapolation was used to calculate the actual values of $\case{\Phi}{\Omega}$ on $\scri^+$. The solid line represents the data as computed during evolution, measuring the evolution by the coordinate time. The time dependence and amplitude of the dotted line have been calculated a posteriori following the indications explained in subsection \ref{bondi}.

An example of the convergence of the rescaled scalar function at null infinity,
obtained evolving the \CZ{} ($C_{Z4c}=1$) system, is shown on the right in figure
\ref{phiscri}. Qualitatively similar convergence plots have been obtained for tens of
similar simulations, with errors for GBSSN typically slightly larger than for Z4. 
For figure \ref{phiscri} we again used extrapolation to calculate the values of $\Phi/\Omega$ at $\scri^+$ and then checked convergence. The expected 4th order convergence is excellent when the differences are significant, but the coincidence is not so good when they are very small, as is expected. 
The error estimate included in the plot is the difference between the estimated exact value of $\Phi/\Omega$ and the highest resolution. 

Regarding the performance of different systems of evolution equations, we find that all the three systems of equations we have used (GBSSN, \CZ{} ($C_{Z4c}=1$) and \CZ{} ($C_{Z4c}=0$)) are appropriate, and none performs clearly better than the others. Accuracy as analyzed in a convergence tests varies between systems and from case to case, and different parameters (damping parameters, dissipation, etc.) may have to be used depending on the evolution system.
As an example, the data in \fref{phiscri} belong to a simulation with \CZ{} ($C_{Z4c}=1$); the same simulation carried out with \CZ{} ($C_{Z4c}=0$) showed a convergence which was practically the same, whereas that of the equivalent GBSSN simulation was considerably less accurate. At the stage when the equations were not stable yet, analyzing instabilities with GBSSN was somewhat easier, as it has one variable less than \CZ{}, but having different evolution systems available for experiments was extremely valuable.

Both the gauge waves and the scalar field evolution are well-behaved for a value of $a$ as large as $a=1000$ using $\kappa_1=0.003$ and 5th order extrapolation for the ghost points 
for the three formulations we have tested, GBSSN, \CZ{} ($C_{Z4c}=0$) and \CZ{} ($C_{Z4c}=1$). With respect to a lower limit of $a$, with $a=0.8$ and $\kappa_1=3$ the GBSSN equations are still stable, but the only example with a value of $a$ below 1 that we could achieve using the \CZ{} equations was with $a=0.88$, $\kappa_1=1$ and only for $C_{Z4c}=1$. 
We have determined the largest stable Courant factor numerically with spatial resolutions of about 400 points using $r_{\!\!\scri}=1$, $a=1$ and setting $\kappa_1=1.5$. We find the maximal allowed Courant factor to be  0.200 for the \CZ{} equations when $C_{Z4c}=0$  (although 0.224 is possible in this case for $\kappa_2=-1$), 0.224 for \CZ{}  with $C_{Z4c}=1$, and 0.252 for the GBSSN  system.

\vspace{-1ex}

\section{Conclusions}

In this paper we considered the hyperboloidal initial value problem for regular asymptotically flat spacetimes, i.e. they do not contain black holes. The data sets we considered do however contain strong gauge wave effects, and large amplitude scalar waves, where a $\sim3\%$ increase in the amplitude of the scalar field would lead to the formation of a black hole.

We derived the equations for the conformal quantities and calculated appropriate initial data. We successfully evolved the Einstein equations in their unconstrained GBSSN and \CZ{} formulations (standard choices in current production codes) on spherically symmetric hyperboloidal foliations, both as gauge waves and coupled to a massless scalar field. 
Obtaining these results was not trivial: we had to use appropriate gauge conditions, find well-behaved evolution variables and take into consideration other aspects such as constraint damping to achieve a stable numerical evolution. 
We have also taken advantage of some parameter freedom and studied their effect on the simulations. Some choices resulted in a clear improvement in the quality of the simulations, whereas in other cases their effect was more difficult to evaluate. 
We have successfully extracted the scalar field signal at $\scri^+$ and checked its convergence, showing the potential of hyperboloidal foliations to be used in more ambitious simulations scenarios such as binary black hole inspirals and mergers. 

Our next step will be an extensive study on gauge conditions, adding the shift to the set of evolved quantities. We will also consider strong field initial data and the collapse of a scalar field into a black hole.

\ack

SH is grateful to  An{\i}l Zengino\u{g}lu for discussions.
AV and SH acknowledge the support of the European Union FEDER founds, the Spanish Ministry of Economy and Competitiveness (FPA2010-16495, CSD2009-00064, FPA2013-41042-P), the FPU-grant AP2010-1697
and the Conselleria d'Economia i Competitivitat of the Govern de les Illes Balears. DH is supported in part by DFG grant SFB/Transregio 7, and gratefully acknowledges the hospitality of UIB. 
Most of the algebraic derivations were performed using the Mathematica package {\tt xAct} \cite{xAct}. 

\appendix

\section{3+1 decomposed equations}\label{app3+1}

Here we present the 3+1 decomposed Einstein equations derived from their 4-dimensional form \eref{einsteineq} and \eref{einsteinten} using the quantities defined in subsection \ref{sub3+1}.
In the following equations $\bar D$ is the covariant derivative associated with $\bar\gamma_{ab}$, the 3-dimensional spatial metric that is also used to raise and lower indices, and we use the notation $\bar\bigtriangleup\equiv\bar\gamma^{ab}\bar D_a\bar D_b$ and $\partial_\perp=\partial_t-\mathcal{L}_\beta$. 
Our 3+1 decomposed evolution equations are:
\begin{subequations}\label{adm}
\begin{eqnarray}
\fl \partial_\perp \bar\gamma_{ab} = & -2\alpha \bar K_{ab}  , \\ 
\fl \partial_\perp \bar K_{ab} = & \alpha\left[R[\bar D]_{ab}-2 \bar K_{a}^{c} \bar K_{bc} + \bar K_{ab} (\bar K-2C_{Z4c}\Theta) + 2 \bar D_{(a}Z_{b)} -  \frac{\kappa_1 (1+\kappa_2) \bar\gamma_{ab} \Theta}{\Omega}  \right] -  \bar D_{b}\bar D_{a}\alpha     \nonumber \\ \fl & 
 + \frac{3 \bar\gamma_{ab} \left[(\partial_\perp\Omega)^2-\alpha^2\bar D^{c}\Omega \bar D_{c}\Omega\right]}{\alpha \Omega^2}   + \frac{4 \alpha Z_{(a} \bar D_{b)}\Omega}{\Omega} + \frac{2 \alpha \bar D_{b}\bar D_{a}\Omega}{\Omega} -  \frac{2 \alpha \bar\gamma_{ab} Z^{c} \bar D_{c}\Omega}{\Omega}\nonumber \\  \fl & 
 + \frac{\bar\gamma_{ab} \bar D^{c}\alpha \bar D_{c}\Omega}{\Omega} + \frac{\alpha \bar\gamma_{ab} \bar\bigtriangleup\Omega}{\Omega} + \frac{2 \bar K_{ab} \partial_\perp\Omega}{\Omega} + \frac{\bar\gamma_{ab} (\bar K-2C_{Z4c}\Theta) \partial_\perp\Omega}{\Omega} \nonumber \\  \fl & 
 + \frac{\bar\gamma_{ab} \partial_\perp\alpha \partial_\perp\Omega}{\alpha^2 \Omega}  -  \frac{\bar\gamma_{ab} \partial_\perp\partial_\perp\Omega}{\alpha \Omega} +4\pi\alpha\left[\bar\gamma_{ab}(S-\rho) - 2 S_{ab}\right], \\
\fl \partial_\perp \Theta = & \frac{\alpha}{2}\left[R[\bar D] - \bar K_{ab} \bar K^{ab} +\bar K(\bar K-2C_{Z4c}\Theta) + 2 \bar D_{a}Z^{a} - \frac{2\kappa_1 (2+\kappa_2) \Theta}{\Omega}\right]  -  C_{Z4c} Z^{a} \bar D_{a}\alpha   \nonumber \\  \fl & 
+ \frac{3 \left[(\partial_\perp\Omega)^2-\alpha^2 \bar D^{a}\Omega \bar D_{a}\Omega\right]}{\alpha \Omega^2} + \frac{2 \alpha \bar\bigtriangleup\Omega}{\Omega}  + \frac{2 (K-2C_{Z4c}\Theta) \partial_\perp\Omega}{\Omega} - 8 \pi \alpha \rho  , \\
\fl \partial_\perp Z_{a} = & \alpha\left[\bar D_{b}\bar K_{a}^{b} - \bar D_{a}\bar K+ \bar D_{a}\Theta - 2 \bar K_{ab} Z^{b} - \frac{\kappa_1 Z_{a}}{\Omega} \right]   -  C_{Z4c} \Theta \bar D_{a}\alpha + \frac{2 \alpha \Theta \bar D_{a}\Omega}{\Omega}   \nonumber \\  \fl & 
 -  \frac{2 \bar D_{a}\partial_\perp\Omega}{\Omega}  -  \frac{2 \alpha \bar K_{a}{}^{b} \bar D_{b}\Omega}{\Omega}  -  \frac{2 Z_{a} \partial_\perp\Omega}{\Omega} + \frac{2 \bar D_{a}\alpha \partial_\perp\Omega}{\alpha \Omega} -8 \pi \alpha J_{a} .
\end{eqnarray}
\end{subequations}
The 3+1 decomposed constraint equations are:
\begin{subequations}
\begin{eqnarray}
\fl \mathcal{H} = R[\bar D] - \bar K_{ab} \bar K^{ab} + \bar K^2  + \frac{6 \left[(\partial_\perp\Omega)^2-\alpha^2\bar D^{a}\Omega \bar D_{a}\Omega\right]}{\alpha^2 \Omega^2} 
 + \frac{4 \bar\bigtriangleup\Omega}{\Omega} + \frac{4 \bar K\partial_\perp\Omega}{\alpha \Omega}  - 16 \pi \rho , \qquad\ \  \\
\fl \mathcal{M}^a =  \bar D_{b}\bar K^{ab} -  \bar\gamma^{ab} \bar D_{b}\bar K-  \frac{2 \bar K^{ab} \bar D_{b}\Omega}{\Omega} -  \frac{2 \bar\gamma^{ab} \bar D_{b}\partial_\perp\Omega}{\alpha \Omega} + \frac{2 \bar\gamma^{ab} \bar D_{b}\alpha \partial_\perp\Omega}{\alpha^2 \Omega} -8 \pi J^{a}  .
\end{eqnarray}
\end{subequations}

The coefficient $C_{Z4c}$ denotes the Z4 non-damping non-principal-part terms that are dropped in the Z4c formulation \cite{Bernuzzi:2009ex,Weyhausen:2011cg}. 
They are kept in the CCZ4 one \cite{Alic:2011gg}, but here we are not able to compare with this last formulation, as there the Ricci scalar in the equation of motion of the trace of the extrinsic curvature is not substituted, while we eliminate it using the evolution equation of $\Theta$. Note that this difference does not play a role if the algebraic constraints are satisfied, because the properties of the continuum partial differential equations are unaffected by a change of variables. By choosing $C_{Z4c}=0$ our \CZ{} equations are the close as possible to the Z4c formulation, whereas setting $C_{Z4c}=1$ maintains all the non-principal-part terms. 

\section{Conformal equations (GBSSN and \CZ{} formulation)}\label{apptens}

In subsection \ref{subconf} the quantities appearing in the GBSSN formulation were introduced: the conformal factor $\chi$, the trace-free part of the extrinsic curvature $A_{ab}$, its trace $K$ and the difference of contracted Christoffel symbols $\Delta\Gamma^a$ and $\Lambda^a$. Here we list the corresponding equations including the terms belonging to the \CZ{} formulation as well as the expression used for the Ricci tensor.  

The spatial Ricci tensor related to $\gamma_{ab}$, expressed using $\Delta\Gamma^a_{bc}$ and $\Delta\Gamma^a$, is \cite{Alcubierre:2010is,Brown:2009dd}
\begin{eqnarray}\label{ricciexpr}
	  R[D]_{ab}  = & -\frac{1}{2} \gamma^{cd} \hat D_c \hat D_d \gamma_{ab} 
	+ \gamma_{c(a}D_{b)}\Delta\Gamma^c - \gamma^{cd} \gamma_{e(a} R[\hat D]_{b)cd}{}^e \nonumber\\
       & + \gamma^{cd} \left( 2 \Delta\Gamma^e_{c(a} \Delta\Gamma_{b)ed} 
	+ \Delta\Gamma^e_{ac} \Delta\Gamma_{ebd} \right)  ,
\end{eqnarray}
where $D$ is the covariant derivative associated with $\gamma_{ab}$ and $\hat D_a$ the one that uses the connection functions $\hat\Gamma^a_{bc}$. The quantity $R[\hat D]^a{}_{bcd}$ is the Riemann tensor built from the background metric $\hat \gamma_{ab}$. 

The contraction of this Ricci tensor $R[D]=\gamma^{ab}R[D]_{ab}$ is to be substituted in the Hamiltonian constraint \eref{tensorH}. However, for the evolution equations \eref{tensorA} and \eref{tensorTheta} we will use 
\begin{eqnarray}\label{riccizexpr}
	  R[D]_{ab} + 2D_{(a}Z_{b)}  = & -\frac{1}{2} \gamma^{cd} \hat D_c \hat D_d \gamma_{ab} 
	+ \gamma_{c(a}D_{b)}\Lambda^c - \gamma^{cd} \gamma_{e(a} R[\hat D]_{b)cd}{}^e \nonumber\\
       & + \gamma^{cd} \left( 2 \Delta\Gamma^e_{c(a} \Delta\Gamma_{b)ed} 
	+ \Delta\Gamma^e_{ac} \Delta\Gamma_{ebd} \right) \ 
\end{eqnarray}
and its corresponding Ricci scalar, as this expression includes the evolution variable $\Lambda^a$. 

In the following equations we use the covariant metric to raise and lower indices (except those of $Z_a$, $J^a$ and $\mathcal{M}^a$, which are moved with $\bar \gamma_{ab}=\case{\gamma_{ab}}{\chi}$) and again the notation $\bigtriangleup\equiv\gamma^{ab} D_a D_b$. The Z4 variable $Z_a$ is eliminated in favour of $\Lambda^a$ according to \eref{tensorZ}. 

The final GBSSN and \CZ{} evolution equations are:
\begin{subequations}\label{tensoreqs}
\begin{eqnarray}
\fl \partial_\perp\chi = & \case{2}{3} \alpha \chi (K+2\Theta) + \case{1}{3} \chi \partial_\perp\ln\gamma
  , \\ 
\fl \partial_\perp\gamma_{ab} = & -2 A_{ab} \alpha + \case{1}{3} \gamma_{ab} \partial_\perp\ln\gamma
  , \\ 
\fl \partial_\perp A_{ab} = & \left[   \alpha \chi \left(R[D]_{ab} + 2D_{(a}Z_{b)} \right) -  \chi D_{a}D_{b}\alpha  - D_{(a}\alpha D_{b)}\chi  -  \frac{\alpha D_{a}\chi D_{b}\chi}{4 \chi} + \case{1}{2} \alpha D_{a}D_{b}\chi \right. \nonumber \\  
\fl & \left. + 2 Z_{(a} \alpha D_{b)}\chi  + \frac{2\alpha D_{(a}\chi D_{b)}\Omega}{\Omega} + \frac{2 \alpha \chi D_{a}D_{b}\Omega}{\Omega}   + \frac{4\alpha \chi  Z_{(a} D_{b)}\Omega}{\Omega}  - 8 \pi \alpha \chi S_{ab} \right]^{TF} \nonumber \\ 
\fl & -2 \alpha A_{a}^{c} A_{bc} + \alpha A_{ab} [K+2(1-C_{Z4c})\Theta] + \frac{2 A_{ab} \partial_\perp\Omega}{\Omega} + \case{1}{3} A_{ab} \partial_\perp\ln\gamma
  , \label{tensorA} \\
\fl \partial_\perp K = & \alpha\left[A_{ab} A^{ab} + \case{1}{3} (K+2\Theta)^2  + \frac{\kappa_1(1-\kappa_2) \Theta}{\Omega}\right] - \chi \bigtriangleup\alpha  + \case{1}{2} D^{a}\alpha D_{a}\chi + 2 C_{Z4c} Z^{a} D_{a}\alpha  \nonumber \\ 
\fl & + \frac{3 [(\partial_\perp\Omega)^2-\alpha^2\chi D^{a}\Omega D_{a}\Omega]}{\Omega^2 \alpha} -  \frac{2 C_{Z4c} \alpha Z^{a} D_{a}\Omega}{\Omega} + \frac{3 \chi D^{a}\alpha D_{a}\Omega}{\Omega} -  \frac{\alpha D^{a}\chi D_{a}\Omega}{2 \Omega}  \nonumber \\ 
\fl & + \frac{\alpha \chi \bigtriangleup\Omega}{\Omega} + \frac{[K+4\Theta] \partial_\perp\Omega}{\Omega} + \frac{3 \partial_\perp\alpha \partial_\perp\Omega}{\Omega \alpha^2} -  \frac{3 \partial_\perp\partial_\perp\Omega}{\Omega \alpha} + 4 \pi \alpha (\rho+S)
  , \\ 
\fl \partial_\perp \Lambda^a = &  \frac{2Z^b \tilde D_{b}\beta^{a}}{\chi} +\alpha\left[ 2 A^{bc} \Delta \Gamma^{a}_{bc} -  \case{4}{3} D^{a}K  -  \case{2}{3} D^{a}\Theta  -  \frac{3 A^{ab} D_{b}\chi}{\chi} -  \frac{4 Z^{a}(K+2\Theta)}{3 \chi} -  \frac{2 \kappa_1 Z^{a} }{\Omega \chi} \right] \nonumber \\ 
\fl &  + \gamma^{bc} \hat D_{b}\hat D_{c}\beta^{a} - \gamma^{bc} R[\hat D]^{a}{}_{bcd} \beta^{d}  - 2 A^{ab} D_{b}\alpha - 2 C_{Z4c} \Theta D^{a}\alpha  -  \frac{4 \alpha A^{ab} D_{b}\Omega}{\Omega} \nonumber \\ 
\fl & -  \frac{2 \alpha (2K+\Theta) D^{a}\Omega}{3 \Omega} +  \frac{2 C_{Z4c} \alpha \Theta D^{a}\Omega}{\Omega}  -  \frac{4 D^{a}\partial_\perp\Omega}{\Omega}  + \frac{4 D^a\alpha \partial_\perp\Omega}{\Omega \alpha}  -  \frac{4 Z^{a} \partial_\perp\Omega}{\Omega \chi} \nonumber \\ 
\fl & -  \case{1}{6} D^a\partial_\perp\ln\gamma -  \case{1}{3} \Delta \Gamma^{a} \partial_\perp\ln\gamma -  \frac{2 Z^{a} \partial_\perp\ln\gamma}{3 \chi} -  \frac{16 \pi J^{a} \alpha}{\chi}
  , \label{tensLambda}\\ 
\fl \partial_\perp \Theta = & \case{\alpha}{2}\left[ \chi (R[D]+ 2 D^{a}Z_{a}) -  A_{ab} A^{ab} + \case{2}{3} (K+2\Theta)^2  - 2 C_{Z4c} \Theta (K+2\Theta)  -  \frac{2\kappa_1(2+\kappa_2) \Theta}{\Omega} \right]  \nonumber \\ 
\fl & + \alpha\bigtriangleup\chi -  \frac{5 \alpha D^{a}\chi D_{a}\chi}{4 \chi}  -  C_{Z4c} Z^{a} D_{a}\alpha -  \frac{C_{Z4c}\alpha Z^{a} D_{a}\chi}{2 \chi} + \frac{2 \alpha \chi \bigtriangleup\Omega}{\Omega}  -  \frac{\alpha D^{a}\chi D_{a}\Omega}{\Omega} \nonumber \\ 
\fl & + \frac{3 [(\partial_\perp\Omega)^2-\alpha^2\chi D^a\Omega D_a\Omega]}{\Omega^2 \alpha}   + \frac{2 K\partial_\perp\Omega}{\Omega} - 8 \pi \alpha \rho
  . \label{tensorTheta} 
\end{eqnarray}
\end{subequations}
The GBSSN and \CZ{} constraint equations are:
\begin{subequations}\label{tensorceqs}
\begin{eqnarray}
\fl \mathcal{H} = & \chi R[D] - A_{ab} A^{ab} + \case{2}{3} (K+2\Theta)^2 + 2 \bigtriangleup\chi -  \frac{5 D^{a}\chi D_{a}\chi}{2 \chi} + \frac{6 [(\partial_\perp\Omega)^2-\alpha^2\chi D^{a}\Omega D_{a}\Omega]}{\Omega^2 \alpha^2}  \nonumber \\ 
\fl &  -  \frac{2  D^{a}\chi D_{a}\Omega}{\Omega} + \frac{4 \chi \bigtriangleup\Omega}{\Omega}  + \frac{4 (K+2\Theta) \partial_\perp\Omega}{\Omega \alpha} - 16 \pi \rho , \label{tensorH} \\ 
\fl \mathcal{M}_{a} = & D_{b}A_{a}^{b}  -  \case{2}{3} D_{a}(K+2\Theta)  -  \frac{3 A_{a}^{b} D_{b}\chi}{2 \chi}  -  \frac{2 A_{a}^{b} D_{b}\Omega}{\Omega} -  \frac{2 (K+2\Theta) D_{a}\Omega}{3 \Omega} -  \frac{2 D_{a}\partial_\perp\Omega}{\Omega \alpha} \nonumber \\ 
\fl & + \frac{2 D_{a}\alpha \partial_\perp\Omega}{\Omega \alpha^2} 
-8 \pi J_{a}  , \\
%
\fl Z_{a} = & \frac{\gamma_{ab}}{2}\left(\Lambda^{b} - \Delta \Gamma^{b}\right) \label{tensorZ}. 
\end{eqnarray}
\end{subequations}

\section{Spherically symmetric equations}\label{appspher}

The spherically symmetric reduction of the GBSSN and \CZ{} equations for the variable ans\"atze in subsection \ref{subspher} is presented here. 
We will use the Lagrangian formulation of the GBSSN equations. Its condition $\partial_t\gamma=0$ translates to substituting $\partial_\perp\ln\gamma=-2D_a\beta^a$ in the tensorial equations \eref{tensoreqs}, see \cite{Brown:2007nt,Brown:2009dd}. 
Here dots denote time derivatives and primes, derivatives with respect to $r$. The coefficients $\xi_\alpha$, $\kappa_1$ and $\kappa_2$ are left unset to experiment and find the appropriate values for the numerical simulations. The transformation for the trace of the extrinsic curvature $K$ into the new variable $\K$, the rescaling of $\Theta$ into $\Te$, and the inclusion of the constraint damping term into $\dot\Lambda^r$ described in section \ref{regularizingsec} are already performed. 
The GBSSN system is obtained by setting the Z4 quantities $\Te$ and $Z_r$ to zero and not evolving $\Te$. For the \CZ{} one, $\Te$ is evolved in time and $Z_r$ is substituted in the RHSs using \eref{Zrdef}.
Note that for simplicity we show the equations with the unrescaled scalar field variables.
The spherically symmetric GBSSN and \CZ{} evolution equations are:
\begin{subequations}
\begin{eqnarray}
 \fl \dot{\chi } =& {\beta^r} \chi ' + \frac{2 \alpha  \chi (\K+2\Te)}{3\Omega}-\frac{{\beta^r} \gamma_{rr}' \chi }{3 \gamma_{rr}}-\frac{2 {\beta^r} \gamma_{\theta\theta}' \chi }{3
   \gamma_{\theta\theta}}   -\frac{4 {\beta^r} \chi }{3 r}-\frac{2 {\beta^r}' \chi }{3} +  \frac{2 {\beta^r} \chi  \Omega '}{\Omega }-\frac{2 \alpha  \chi }{ a  \Omega }, \\ 
 \fl \dot{\gamma_{rr}} =& \frac{2 {\beta^r} \gamma_{rr}'}{3} -2 A_{rr} \alpha -\frac{2 \gamma_{rr} {\beta^r} \gamma_{\theta\theta}'}{3 \gamma_{\theta\theta}} +  \frac{4
   \gamma_{rr} {\beta^r}'}{3}-\frac{4 \gamma_{rr} {\beta^r}}{3 r}   , \\ 
 \fl \dot{\gamma_{\theta\theta}} = & \frac{{\beta^r} \gamma_{\theta\theta}'}{3} + \frac{A_{rr} \gamma_{\theta\theta} \alpha }{\gamma_{rr}}-\frac{\gamma_{\theta\theta} {\beta^r} \gamma_{rr}'}{3 \gamma_{rr}} +  \frac{2 \gamma_{\theta\theta} {\beta^r}}{3 r}-\frac{2 \gamma_{\theta\theta} {\beta^r}'}{3}  , \\ 
 \fl \dot{A_{rr}} = &  {\beta^r} A_{rr}'+\frac{2}{3} \gamma_{rr} \alpha  \chi  {\Lambda^r}'-\frac{\alpha  \chi  \gamma_{rr}''}{3 \gamma_{rr}}+\frac{\alpha  \chi  \gamma_{\theta\theta}''}{3 \gamma_{\theta\theta}}-\frac{2 \chi  \alpha ''}{3}+\frac{\alpha  \chi ''}{3} + \frac{\alpha A_{rr}\left[\K +2(1-C_{Z4c}) \Te\right]}{\Omega} \nonumber \\ 
  \fl &  -\frac{2 \alpha  A_{rr}^2}{\gamma_{rr}}+\frac{4 {\beta^r}' A_{rr}}{3}-\frac{4 {\beta^r} A_{rr}}{3 r}-\frac{{\beta^r} \gamma_{rr}' A_{rr}}{3 \gamma_{rr}}-\frac{2 {\beta^r} \gamma_{\theta\theta}' A_{rr}}{3 \gamma_{\theta\theta}}+\frac{\alpha  \chi  \left(\gamma_{rr}'\right)^2}{2 \gamma_{rr}^2}-\frac{2 \alpha  \chi  \left(\gamma_{\theta\theta}'\right)^2}{3 \gamma_{\theta\theta}^2} \nonumber \\ 
  \fl &-\frac{\alpha \left(\chi '\right)^2}{6 \chi }+\frac{2 \gamma_{rr} \alpha  \chi }{\gamma_{\theta\theta} r^2}-\frac{2 \alpha  \chi }{r^2}-\frac{2 \alpha  {\Lambda^r} \chi \gamma_{rr} }{3 r}+\frac{\alpha  {\Lambda^r} \chi  \gamma_{rr}'}{3}-\frac{\alpha  {\Lambda^r} \chi \gamma_{rr}  \gamma_{\theta\theta}'}{3 \gamma_{\theta\theta}}+\frac{2 \alpha  \chi \gamma_{rr}  \gamma_{\theta\theta}'}{\gamma_{\theta\theta}^2 r}\nonumber \\ 
  \fl &-\frac{2 \alpha  \chi  \gamma_{rr}'}{3 \gamma_{\theta\theta} r} -\frac{4 \alpha  \chi  \gamma_{\theta\theta}'}{3 \gamma_{\theta\theta} r}+\frac{2 \chi  \alpha '}{3 r}+\frac{\chi  \gamma_{rr}' \alpha '}{3 \gamma_{rr}}+\frac{\chi  \gamma_{\theta\theta}' \alpha '}{3 \gamma_{\theta\theta}}-\frac{\alpha  \chi '}{3r}-\frac{\alpha  \gamma_{rr}' \chi '}{6 \gamma_{rr}}-\frac{\alpha  \gamma_{\theta\theta}' \chi '}{6 \gamma_{\theta\theta}}-\frac{2 \alpha ' \chi '}{3} \nonumber \\ 
  \fl &  +\frac{4}{3} Z_r \alpha  \chi '  -\frac{2 \alpha  \chi  \gamma_{rr}' \Omega '}{3 \gamma_{rr} \Omega }-\frac{2 \alpha  \chi  \gamma_{\theta\theta}' \Omega'}{3 \gamma_{\theta\theta} \Omega }+\frac{4 \alpha  \chi ' \Omega '}{3 \Omega }-\frac{4 \alpha  \chi  \Omega '}{3 r \Omega }+\frac{A_{rr} {\beta^r} \Omega '}{\Omega } -\frac{3 \alpha  A_{rr}}{ a  \Omega } \nonumber \\ 
  \fl &+\frac{4 \alpha  \chi  \Omega ''}{3 \Omega } +\frac{8 Z_r \alpha  \chi  \Omega '}{3 \Omega }  -\frac{16}{3} \pi  \alpha  \chi \left(\Phi '\right)^2  , \\ 
 \fl \dot{\K} = &  {\beta^r} \K'-\frac{\chi  \alpha ''\Omega}{\gamma_{rr}} +\frac{\alpha (\K+2\Te)^2}{3\Omega} +  \frac{3 \alpha  A_{rr}^2\Omega}{2 \gamma_{rr}^2}-\frac{2 \chi  \alpha '\Omega}{\gamma_{rr} r}+\frac{\chi \gamma_{rr}' \alpha '\Omega}{2 \gamma_{rr}^2}-\frac{\chi  \gamma_{\theta\theta}' \alpha '\Omega}{\gamma_{rr} \gamma_{\theta\theta}}+\frac{\alpha ' \chi '\Omega}{2 \gamma_{rr}} \nonumber \\ 
  \fl &  +\frac{\kappa_1(1-\kappa_2) \alpha  \Te }{\Omega } +\frac{2 C_{Z4c} Z_r \chi  \alpha '\Omega}{\gamma_{rr}} 
  -\frac{\alpha  \chi  \gamma_{rr}' \Omega '}{2 \gamma_{rr}^2 }+\frac{\alpha  \chi  \gamma_{\theta\theta}' \Omega '}{\gamma_{rr} \gamma_{\theta\theta} }+\frac{3 \chi  \alpha ' \Omega '}{\gamma_{rr}  }   -\frac{\alpha  \chi ' \Omega '}{2 \gamma_{rr} } \nonumber \\ 
  \fl &  -\frac{2 \alpha (\K+2\Te)}{ a \Omega }+\frac{2 \alpha  \chi  \Omega '}{\gamma_{rr} r }+\frac{\alpha  \chi  \Omega ''}{\gamma_{rr} } -\frac{2C_{Z4c} Z_r \alpha  \chi  \Omega '}{\gamma_{rr} } +\frac{3 \alpha }{ a ^2 \Omega} -\frac{3 \alpha  \chi \left(\Omega '\right)^2}{\gamma_{rr} \Omega} \nonumber \\ 
  \fl & +\frac{8\pi(\Pi-{\beta^r}\Phi ')^2\Omega}{\alpha}  , \\ 
 \fl \dot{{\Lambda^r}} = & {\beta^r} {\Lambda^r}' -\frac{2 \alpha (2 \K'+\Te')}{3 \gamma_{rr}\Omega}+\frac{{\beta^r} \gamma_{rr}''}{6 \gamma_{rr}^2}+\frac{{\beta^r} \gamma_{\theta\theta}''}{3 \gamma_{rr} \gamma_{\theta\theta}} +\frac{2 A_{rr} \alpha }{\gamma_{rr}^2 r}  -\frac{2 A_{rr} \alpha }{\gamma_{rr} \gamma_{\theta\theta} r}  -\frac{2 A_{rr} \alpha '}{\gamma_{rr}^2} -\frac{3 A_{rr} \alpha  \chi '}{\gamma_{rr}^2 \chi } \nonumber \\ 
  \fl & +\frac{A_{rr} \alpha  \gamma_{rr}'}{\gamma_{rr}^3}+\frac{A_{rr} \alpha  \gamma_{\theta\theta}'}{\gamma_{rr}^2 \gamma_{\theta\theta}}  -\frac{{\beta^r} \left(\gamma_{\theta\theta}'\right)^2}{\gamma_{rr} \gamma_{\theta\theta}^2} +\frac{2{\beta^r} \gamma_{rr}'}{3 \gamma_{rr} \gamma_{\theta\theta} r}  -\frac{8 {\beta^r} \gamma_{\theta\theta}'}{3 \gamma_{rr} \gamma_{\theta\theta} r}+\frac{4 {\beta^r} \gamma_{\theta\theta}'}{3 \gamma_{\theta\theta}^2 r}+\frac{2 \gamma_{\theta\theta}' {\beta^r}'}{3 \gamma_{rr} \gamma_{\theta\theta}}  \nonumber \\ 
  \fl &-\frac{10 {\beta^r}}{3 \gamma_{rr} r^2} +\frac{2 {\beta^r}}{3 \gamma_{\theta\theta} r^2}+\frac{4 {\beta^r}'}{3 \gamma_{rr} r} +\frac{4 {\beta^r}'}{3 \gamma_{\theta\theta} r}+\frac{4 {\beta^r}''}{3 \gamma_{rr}}  -\frac{4 (\K+2\Te) Z_r \alpha }{3 \gamma_{rr}\Omega}+\frac{2Z_r {\beta^r} \gamma_{rr}'}{3 \gamma_{rr}^2} \nonumber \\ 
  \fl & +\frac{4 Z_r {\beta^r} \gamma_{\theta\theta}'}{3 \gamma_{rr} \gamma_{\theta\theta}} +\frac{8 Z_r {\beta^r}}{3 \gamma_{rr} r}+\frac{4 Z_r {\beta^r}'}{3 \gamma_{rr}}-\frac{2 C_{Z4c} \Te  \alpha '}{\gamma_{rr}\Omega}  +\frac{2 C_{Z4c} \alpha \Te \Omega '}{\gamma_{rr} \Omega^2 } -\frac{4 A_{rr} \Omega ' \alpha }{\gamma_{rr}^2 \Omega }+\frac{4 Z_r \alpha }{ a  \gamma_{rr} \Omega } \nonumber \\ 
  \fl & -\frac{2 \kappa_1 \left(\frac{1}{2}\gamma_{rr}\Lambda^r+\frac{1}{r}-\frac{\gamma_{rr}}{\gamma_{\theta\theta}r}-\frac{\gamma_{rr}'}{4\gamma_{rr}}+\frac{\gamma_{\theta\theta}'}{2\gamma_{\theta\theta}}\right) \alpha }{\gamma_{rr} \Omega } + \frac{16 \pi  (\Pi-{\beta^r}\Phi ')  \Phi '}{\gamma_{rr} }   , \\ 
 \fl \dot{\alpha } = & {\beta^r} \alpha ' -\frac{\K \alpha ^2}{\Omega} -\xi_{\alpha}\left[\alpha^2-\left(\frac{\sqrt{r^2+a^2\Omega^2}}{a}\right)^2\right]  -\frac{3 \alpha  {\beta^r} \Omega '}{\Omega } + \frac{2 {\beta^r} \sqrt{r^2+ a ^2 \Omega ^2} \Omega '}{ a  \Omega }  -\frac{r{\beta^r} }{ a^2\Omega}  , \\ 
 \fl \dot{\Phi } = &  \Pi    , \\ 
 \fl \dot{\Pi } = & 2 {\beta^r} \Pi '-{\beta^r}^2 \Phi ''  + \frac{\chi  \Phi '' \alpha ^2}{\gamma_{rr}}+ \frac{2 \chi  \Phi ' \alpha ^2}{\gamma_{rr} r}-\frac{\chi  \gamma_{rr}' \Phi '
   \alpha ^2}{2 \gamma_{rr}^2}+\frac{\chi  \gamma_{\theta\theta}' \Phi ' \alpha ^2}{\gamma_{rr} \gamma_{\theta\theta}}-\frac{\Phi ' \chi ' \alpha ^2}{2 \gamma_{rr}}-\frac{2 \chi  \Phi '
   \Omega ' \alpha ^2}{\gamma_{rr} \Omega } \nonumber \\ 
  \fl &   +\frac{\alpha (\K+2\Te) (\Pi-{\beta^r} \Phi ')}{\Omega} +\frac{\chi  \alpha ' \Phi ' \alpha }{\gamma_{rr}}-{\beta^r} {\beta^r}' \Phi '+\frac{(\dot{\alpha} -{\beta^r} \alpha ' )(\Pi-{\beta^r} \Phi ') }{\alpha } \nonumber \\ 
  \fl & +\frac{ (\Pi-\beta^r\Phi')  {\beta^r}\Omega ' }{\Omega }-\frac{3 \alpha  (\Pi-\beta^r\Phi') }{ a  \Omega }  , \\ 
 \fl \dot{\Te } = &  {\beta^r} \Te '+\frac{1}{2} \alpha  \chi  {\Lambda^r}'\Omega-\frac{\alpha  \chi  \gamma_{rr}''\Omega}{4 \gamma_{rr}^2}-\frac{\alpha  \chi  \gamma_{\theta\theta}''\Omega}{2\gamma_{rr} \gamma_{\theta\theta}}+\frac{\alpha  \chi ''\Omega}{\gamma_{rr}}  +\frac{ \alpha (\K+2\Te)^2}{3\Omega}  -\frac{C_{Z4c} \alpha  \Te (\K+2\Te)}{\Omega}  \nonumber \\ 
  \fl & -\frac{C_{Z4c} Z_r \chi  \alpha '\Omega}{\gamma_{rr}}-\frac{C_{Z4c} Z_r \alpha  \chi '\Omega}{2 \gamma_{rr}} -\frac{\kappa_1(2+\kappa_2) \alpha  \Te }{\Omega } -\frac{3 \alpha  A_{rr}^2\Omega}{4 \gamma_{rr}^2} +\frac{\alpha  {\Lambda^r} \chi \Omega}{r}+\frac{3 \alpha  \chi  \left(\gamma_{rr}'\right)^2\Omega}{8 \gamma_{rr}^3}\nonumber \\ 
  \fl &+\frac{\alpha  \chi  \left(\gamma_{\theta\theta}'\right)^2\Omega}{4 \gamma_{rr} \gamma_{\theta\theta}^2}-\frac{5 \alpha  \left(\chi '\right)^2\Omega}{4 \gamma_{rr} \chi }-\frac{\alpha  \chi  \gamma_{rr}'\Omega}{2 \gamma_{rr} \gamma_{\theta\theta} r}+\frac{\alpha  {\Lambda^r} \chi  \gamma_{rr}'\Omega}{4\gamma_{rr}}-\frac{\alpha  \chi  \gamma_{\theta\theta}'\Omega}{\gamma_{rr} \gamma_{\theta\theta} r}+\frac{\alpha  {\Lambda^r} \chi  \gamma_{\theta\theta}'\Omega}{2 \gamma_{\theta\theta}}+\frac{2 \alpha  \chi '\Omega}{\gamma_{rr} r}  \nonumber \\ 
  \fl & -\frac{\alpha  \gamma_{rr}' \chi '\Omega}{2 \gamma_{rr}^2}+\frac{\alpha  \gamma_{\theta\theta}' \chi '\Omega}{\gamma_{rr} \gamma_{\theta\theta}} -\frac{\alpha \chi  \gamma_{rr}' \Omega '}{\gamma_{rr}^2  }+\frac{2 \alpha  \chi  \gamma_{\theta\theta}' \Omega '}{\gamma_{rr} \gamma_{\theta\theta} }-\frac{\alpha  \chi ' \Omega '}{\gamma_{rr} }+\frac{4 \alpha  \chi  \Omega '}{\gamma_{rr} r } +\frac{2 \alpha  \chi  \Omega ''}{\gamma_{rr} }\nonumber \\ 
  \fl &-\frac{2 \alpha (\K+2\Te)}{ a  \Omega }+\frac{3 C_{Z4c} \alpha  \Te }{ a  \Omega }+\frac{3 \alpha }{ a ^2 \Omega }-\frac{3 \alpha  \chi  \left(\Omega '\right)^2}{\gamma_{rr} \Omega } -4\pi\Omega\left[ \frac{(\Pi-{\beta^r} \Phi ')^2}{\alpha} +\frac{ \alpha  \chi  \left(\Phi '\right)^2}{\gamma_{rr}} \right]. \qquad\ 
\end{eqnarray}
\end{subequations}

The spherically symmetric GBSSN and \CZ{}  constraint equations read:
\begin{subequations}
\begin{eqnarray}
 \fl \mathcal{H} = &   -\frac{3 A_{rr}^2}{2 \gamma_{rr}^2}+\frac{2(\K+2\Te)^2}{3\Omega^2}+\frac{\chi  \left(\gamma_{\theta\theta}'\right)^2}{2 \gamma_{rr} \gamma_{\theta\theta}^2}-\frac{5 \left(\chi '\right)^2}{2
   \gamma_{rr} \chi }-\frac{2 \chi }{\gamma_{rr} r^2}+\frac{2 \chi }{\gamma_{\theta\theta} r^2}+\frac{2 \chi  \gamma_{rr}'}{\gamma_{rr}^2 r}-\frac{6 \chi  \gamma_{\theta\theta}'}{\gamma_{rr} \gamma_{\theta\theta} r} \nonumber \\ 
  \fl &+\frac{\chi  \gamma_{rr}' \gamma_{\theta\theta}'}{\gamma_{rr}^2 \gamma_{\theta\theta}}-\frac{\gamma_{rr}' \chi '}{\gamma_{rr}^2}+\frac{2 \gamma_{\theta\theta}' \chi '}{\gamma_{rr} \gamma_{\theta\theta}}+\frac{4 \chi '}{\gamma_{rr} r}-\frac{2 \chi  \gamma_{\theta\theta}''}{\gamma_{rr} \gamma_{\theta\theta}}+\frac{2 \chi ''}{\gamma_{rr}} -\frac{2 \chi  \gamma_{rr}' \Omega '}{\gamma_{rr}^2 \Omega }+\frac{4 \chi 
   \gamma_{\theta\theta}' \Omega '}{\gamma_{rr} \gamma_{\theta\theta} \Omega } -\frac{2 \chi ' \Omega '}{\gamma_{rr} \Omega } \nonumber \\ 
  \fl &  -\frac{6 \chi  \left(\Omega '\right)^2}{\gamma_{rr} \Omega ^2}+\frac{8 \chi  \Omega '}{\gamma_{rr} r \Omega }+\frac{4 \chi 
   \Omega ''}{\gamma_{rr} \Omega }-\frac{4 (\K+2\Te)}{ a  \Omega^2 }+\frac{6}{ a ^2 \Omega ^2} -8\pi\left[ \frac{(\Pi-{\beta^r}\Phi ')^2}{\alpha^2}+\frac{\chi\left(\Phi '\right)^2}{\gamma_{rr}} \right]   , \qquad\  \\
 \fl \mathcal{M}_r = &   -\frac{\gamma_{rr}' A_{rr}}{\gamma_{rr}^2}+\frac{3 \gamma_{\theta\theta}' A_{rr}}{2 \gamma_{rr} \gamma_{\theta\theta}}-\frac{3 \chi ' A_{rr}}{2 \gamma_{rr} \chi
   }+\frac{3 A_{rr}}{\gamma_{rr} r}+\frac{A_{rr}'}{\gamma_{rr}}-\frac{2 (\K'+2\Te')}{3\Omega} -\frac{2 A_{rr} \Omega '}{\gamma_{rr} \Omega } \nonumber \\ 
  \fl &  + \frac{8 \pi  (\Pi-{\beta^r}\Phi ')  \Phi '}{\alpha }, \\
%
\fl Z_r = & \frac{1}{2}\gamma_{rr}\Lambda^r+\frac{1}{r}-\frac{\gamma_{rr}}{\gamma_{\theta\theta}r}-\frac{\gamma_{rr}'}{4\gamma_{rr}}+\frac{\gamma_{\theta\theta}'}{2\gamma_{\theta\theta}}. \label{Zrdef}
\end{eqnarray}
\end{subequations}

\section*{References}
\bibliographystyle{unsrt}
\bibliography{hypcomp}

\end{document}